\documentclass[aip,graphicx,rsi]{revtex4-1}

\usepackage{graphicx}
\usepackage{dcolumn}
\usepackage{bm}

\usepackage[utf8]{inputenc}
\usepackage[T1]{fontenc}
\usepackage{mathptmx}
\usepackage{etoolbox}

\begin{document}


\title{How does the limited resolution of space plasma analyzers affect the accuracy of space plasma measurements?} 



\author{G. Nicolaou}
\email[]{g.nicolaou@ucl.ac.uk}
\affiliation{Department of Space and Climate Physics, Mullard Space Science Laboratory, University College London, Dorking, Surrey, RH5 6NT, UK}
\author{C. Ioannou}
\affiliation{Department of Space and Climate Physics, Mullard Space Science Laboratory, University College London, Dorking, Surrey, RH5 6NT, UK}

\author{C.J. Owen}
\affiliation{Department of Space and Climate Physics, Mullard Space Science Laboratory, University College London, Dorking, Surrey, RH5 6NT, UK}

\author{D. Verscharen}
\affiliation{Department of Space and Climate Physics, Mullard Space Science Laboratory, University College London, Dorking, Surrey, RH5 6NT, UK}

\author{A. Fedorov}
\affiliation{Institut de Recherche en Astrophysique et Planétologie, 9, Avenue du Colonel ROCHE, BP 4346, 31028 Toulouse Cedex 4, France}

\author{P. Louarn}
\affiliation{Institut de Recherche en Astrophysique et Planétologie, 9, Avenue du Colonel ROCHE, BP 4346, 31028 Toulouse Cedex 4, France}


\date{\today}

\begin{abstract}
We investigate the systematic errors in measured plasma velocity distribution functions and their corresponding velocity moments, arising  from the limited energy and angular resolution of top-hat electrostatic analyzers. For this purpose, we develop a forward model of a concept analyzer that simulates observations of typical solar wind proton plasma particles with their velocities following a Maxwell distribution function. We then review the standard conversion of the observations to physical parameters and evaluate the errors arising from the limited resolution of the modeled instrument. We show that the limited resolution of the instrument results in velocity distributions that underestimate the core and overestimate the tails of the actual Maxwellian plasma velocity distribution functions. As a consequence, the velocity moments of the observed plasma underestimate the proton density and overestimate the proton temperature. Moreover, we show that the examined errors become significant for cold and fast plasma protons. We finally determine a mathematical formula that predicts these systematic inaccuracies based on specific plasma inputs and instrument features. Our results inform and contextualize future evaluations of observations by analyzers in various plasma regimes.  
\end{abstract}


\maketitle 

\section{Introduction}
\label{sec:Introduction}

Top-hat electrostatic analyzers (ESAs) with aperture deflectors and position-sensitive detectors measure the number of incoming charged plasma particles in discrete energy-per-charge, elevation, and azimuth bins.\citep{Young2004,Barabash2006,Nilsson2007,Pollock2016,McComas2017,Livi2022,Owen2020} With these measurements we can construct the three-dimensional (3D) velocity distribution functions (VDFs) of the plasma species measured by the instrument. However, plasma measurements are subject to several errors which propagate inaccuracies to the determined VDFs and determined data products, such as the density, bulk speed, and temperature of the detected species.

For instance, like any other counting experiment, the number of detected particles has a statistical uncertainty governed by Poisson statistics. This uncertainty propagates statistical errors to the physical parameters we determine from the observations.\citep[e.g.][]{Bevington1992,Wilson2015,Nicolaou2014b} Moreover, the statistical uncertainties of the observations lead to systematic uncertainties in the plasma parameters if the typical chi-squared minimization method is used to infer the underlying plasma VDFs.\citep{Stoneking1997,Nicolaou2020,Nicolaou2024rasti} Such systematic errors may lead to artificial correlations between the plasma parameters, which is not only preventing the resolution of physical mechanisms in space, but it may alter the outcome of scientific studies leading to erroneous conclusions. \citep{Nicolaou2024apj}

Plasma particle observations are subject to background noise caused by the instrument electronics. The analysis of  the VDFs constructed from the noisy observations, leads to an overestimation of the zeroth and second order velocity moments which determine the plasma density and temperature, respectively.\citep{Nicolaou2023,Wang2024} Moreover, the background noise affects the determination of particle distribution functions, even when determined by the chi-squared minimization technique.\citep{Nicolaou2022noise}. Therefore, the noise should either be monitored on-board \citep[e.g.][]{McComas2004}, or estimated by on-ground analyses \citep[e.g.][]{Wang2024,Parent2024} and subtracted from observations prior any further analysis.

Other studies have also evaluated the systematic uncertainties in the plasma parameters resulting from non-resolved time variations of the plasma. \citep{Verscharen2011,Nicolaou2019Turbulent} Plasma bulk velocity fluctuations on time-scales below the time-resolution of plasma instruments are expected to result in a broadening of the resolved plasma VDFs and thus, in an overestimation of the plasma temperature. If the velocity fluctuations are more dominant in either the perpendicular or the parallel direction with respect to the background magnetic field, the analysis of the observations may determine false temperature anisotropies. \citep[][]{Verscharen2011}

We also expect systematic uncertainties in the recovered VDFs if the instrument is not capable of resolving VDFs of different species. For instance, solar wind proton VDFs may have significant energy overlap with the VDFs of $\alpha$ particles. In these cases, the analysis may fail to examine the VDFs of the two species separately and return false results.\citep[e.g.][]{Nicolaou2014b,Nicolaou_2018} More specifically, if the alpha particles are treated as protons, the analysis overestimates the actual proton density, speed, and temperature \citep[][]{Zhang2024}.      

Besides the errors mentioned above, we expect additional systematic errors in the plasma interpretations due to the limited angular and energy resolution of ESAs. ESAs sample the plasma particles in discrete energy-per-charge and angular bins, with each bin covering a finite volume in velocity space. The measurements however, cannot resolve the shape of the VDFs within each bin. Although such systematic errors have been discussed in previous publications \citep[e.g.][]{Roberts2021,Wilson2022}, we argue that since there is a significant number of studies using plasma observations by ESAs, there is a need for a dedicated study to provide a detailed methodology to evaluate and estimate them.  

In Section \ref{sec:motivation}, we explain the motivation for this study in detail. Section \ref{sec:methodology} shows the methodology we follow to simulate plasma observations and how we construct the velocity distributions of the plasma. We further explain how we quantify the systematic errors by comparing the constructed distributions and their velocity moments with their respective simulated plasma distributions and their moments. In Sec. \ref{sec:results}, we present our results considering a wide range of plasma proton properties. In Sec. \ref{sec:discussion}, we discuss our results in detail, including the potential impact of the demonstrated uncertainties to scientific studies. We also compare the systematic errors to an analytical function to predict the systematic uncertainties as functions of the plasma VDF derivatives and the instrument resolution. Finally, we discuss a potential mitigation strategy.

\section{Motivation}
\label{sec:motivation}
\begin{figure}
    \centering
    \includegraphics[width=1.0\linewidth]{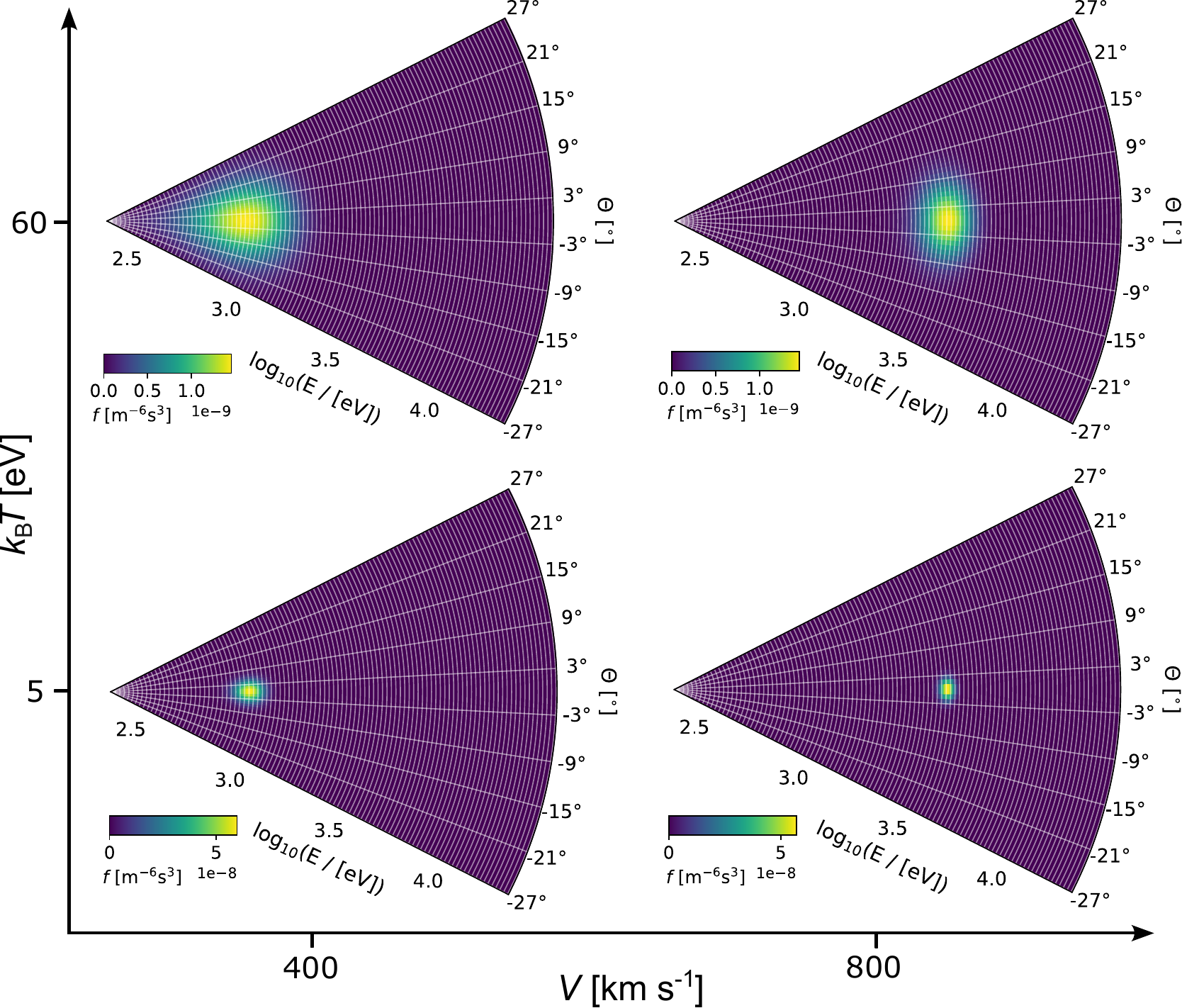}
    \caption{Energy (velocity) distribution function models, for different plasma bulk speeds and temperatures. Each panel shows a modeled distribution as a function of particle energy and elevation direction, for the azimuth direction of the bulk velocity. The white grid on each panel represents the energy and elevation bins of our concept instrument (see Section \ref{sec:concept}).}
    \label{fig:distributions}
\end{figure}

Due to their finite angular and energy resolution, plasma analyzers cannot provide any information about the "shape" of the plasma distribution function within each energy-per-charge and angular bin. Instead, analyzers return one value (number of counts) per bin, which we usually consider as representative of the value of the distribution function at the central energy-per-charge, elevation, and azimuth of the corresponding bin. Analyses of these observations then determine the physical parameters of the plasma. Such simplifications which neglect the details of the instrument response and the shape of the plasma velocity distribution on small, sub-bin scales, may be valid in numerous cases. Here, however, we argue that it is not always safe to adopt them. In Fig. \ref{fig:distributions}, we show modeled Maxwellian energy distribution functions of protons for four different combinations of plasma bulk speeds and temperatures. All four distributions have a bulk velocity vector along elevation angle $\Theta=0^{\circ}$ and azimuth angle $\Phi=0^{\circ}$, and we show two-dimensional 2D "cuts" of the distribution at azimuth $\Phi=0^{\circ}$, as functions of particle energy $E$ and elevation $\Theta$ (see Sec. \ref{sec:methodology} for details). The white grid on each panel shows energy-elevation bins with size $\delta E/E\times \delta\Theta\approx0.05\times 6^{\circ}$. The gradients of the distribution over individual instrument bins become significant as the bulk speed increases and the plasma temperature decreases. Thus, for certain plasma conditions, simplifying the analysis by assuming that the distribution does not vary significantly within each bin can be inappropriate. This study investigates the accuracy of this simplification when applied to standard solar wind proton plasma measurements by an electrostatic analyzer concept and demonstrates the methodology for carrying out accuracy tests. Although the study is carried out using a specific instrument model, the demonstrated methodology can be used for any similar instrument after the proper adjustment of the model.

\section{Methodology}
\label{sec:methodology}
\subsection{Concept Instrument}
\label{sec:concept}
\begin{figure}
    \centering
    \includegraphics[width=0.75\linewidth]{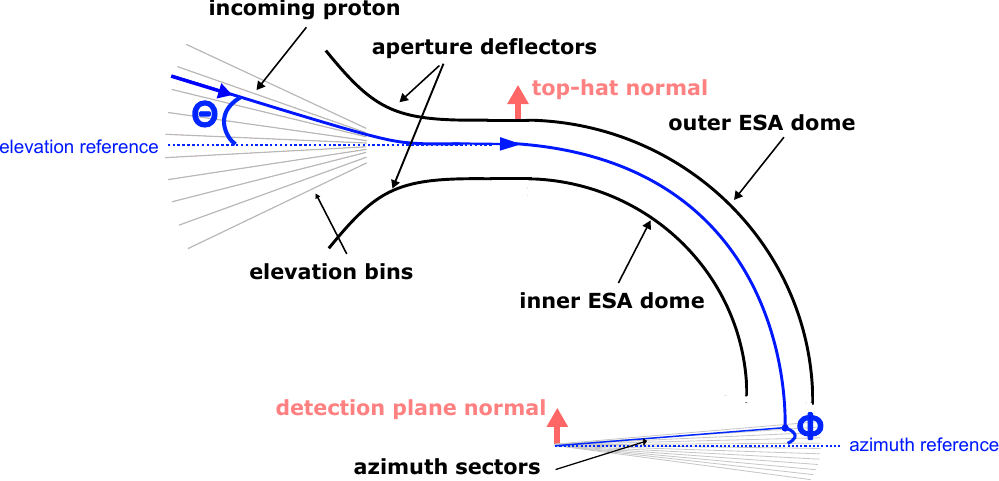}
    \caption{Schematic of our concept instrument. We consider a typical top-hat electrostatic analyzer with aperture deflectors and a position sensitive detector, which can resolve energies, elevation, and azimuth directions of solar wind protons. }
    \label{fig:instrument_diagram}
\end{figure}

We model the response of a typical top-hat electrostatic analyzer for solar wind proton measurements. A diagram of this design is shown in Fig. \ref{fig:instrument_diagram}. In one full acquisition, our concept instrument measures the number of particles in 96 energy-per-charge bins, $E/q$, nine elevation bins, $\Theta$, and eleven azimuth sectors, $\Phi$. The elevation angle is determined as the angle between the velocity vector of the incoming particles and the top-hat plane, while the azimuth angle is the angle between the projection of the particle velocity vector on the top-hat plane (same as the detection plane) and a reference axis onto that plane (see Fig. \ref{fig:instrument_diagram}). Since we simulate protons (charge $q$ = 1), we refer to $E/q$ steps as energy steps $E$ throughout this paper. The 96 $E$ steps are exponentially spaced over a range spanning from 200 eV to 20 keV. The nine $\Theta$ bins sample particles with elevation angles from -24$^{\circ}$ to +24$^{\circ}$, while the eleven $\Phi$ sectors cover azimuth directions from -32$^{\circ}$ to +32$^{\circ}$. The elevation and azimuth bins are equally spaced across their corresponding sampling range of angles. The values of $E$, $\Theta$, and $\Phi$ bins we report above, correspond to the energies, elevations, and azimuths sampled in the center of each bin.

\subsection{Input velocity distribution functions}
\label{sec:inputf}
In order to simulate observations of our concept instrument, we first set-up a velocity distribution function of the “measured” plasma particles. We consider solar wind protons with their velocities following the 3D isotropic Maxwellian distribution function:

\begin{equation}
f(\vec{V})=N_{\mathrm{in}} \left(   \frac{m}{2\pi k_{\mathrm{B}}T_{\mathrm{in}}}   \right)^{\frac{3}{2}}e^{-\frac{m(\vec{V}-\vec{V}_{\mathrm{in}})^2}{2k_{\mathrm{B}}T_{\mathrm{in}}}},
\label{eq:maxwell}
\end{equation}

\noindent where $m$ is the proton mass, $k_{\mathrm{B}}$ is the Boltzmann constant, $\vec{V}$ is the individual proton particle velocity, and $N_{\mathrm{in}}$, $T_{\mathrm{in}}$, and $\vec{V}_{\mathrm{in}}$ are the proton plasma density, temperature, and bulk velocity, respectively. Since electrostatic analyzers resolve particle distributions in a spherical reference frame, we express $f(\vec{V})$ in terms of the individual particle energy $\varepsilon=\frac{1}{2}m \vec{V}\cdot\vec{V}$, elevation $\theta$, and azimuth $\phi$ directions as:

\begin{equation}
f(\varepsilon,\theta,\phi)=N_{\mathrm{in}} \left(   \frac{m}{2\pi k_{\mathrm{B}}T_{\mathrm{in}}}   \right)^{\frac{3}{2}}e^{-\frac{\varepsilon+\varepsilon_{0}-2\sqrt{\varepsilon \varepsilon_{0}}\cos{\omega(\theta,\phi)}}{k_{\mathrm{B}}T_{\mathrm{in}}}},
\label{eq:maxwell_energy}
\end{equation}
\noindent where $\varepsilon_{0}=\frac{1}{2}m\vec{V}_{\mathrm{in}}\cdot\vec{V}_{\mathrm{in}}$ is the bulk energy of the plasma particles and $\omega(\theta,\phi)$ is the angle between the individual particle velocity vector $\vec{V}$ and the bulk velocity vector $\vec{V}_{\mathrm{in}}$. \citep{Nicolaou_2018,Elliott2016,Nicolaou_Livadiotis2016}

\subsection{Forward modeling}
\label{sec:forward_model}
\begin{figure}
    \centering
    \includegraphics[width=1\linewidth]{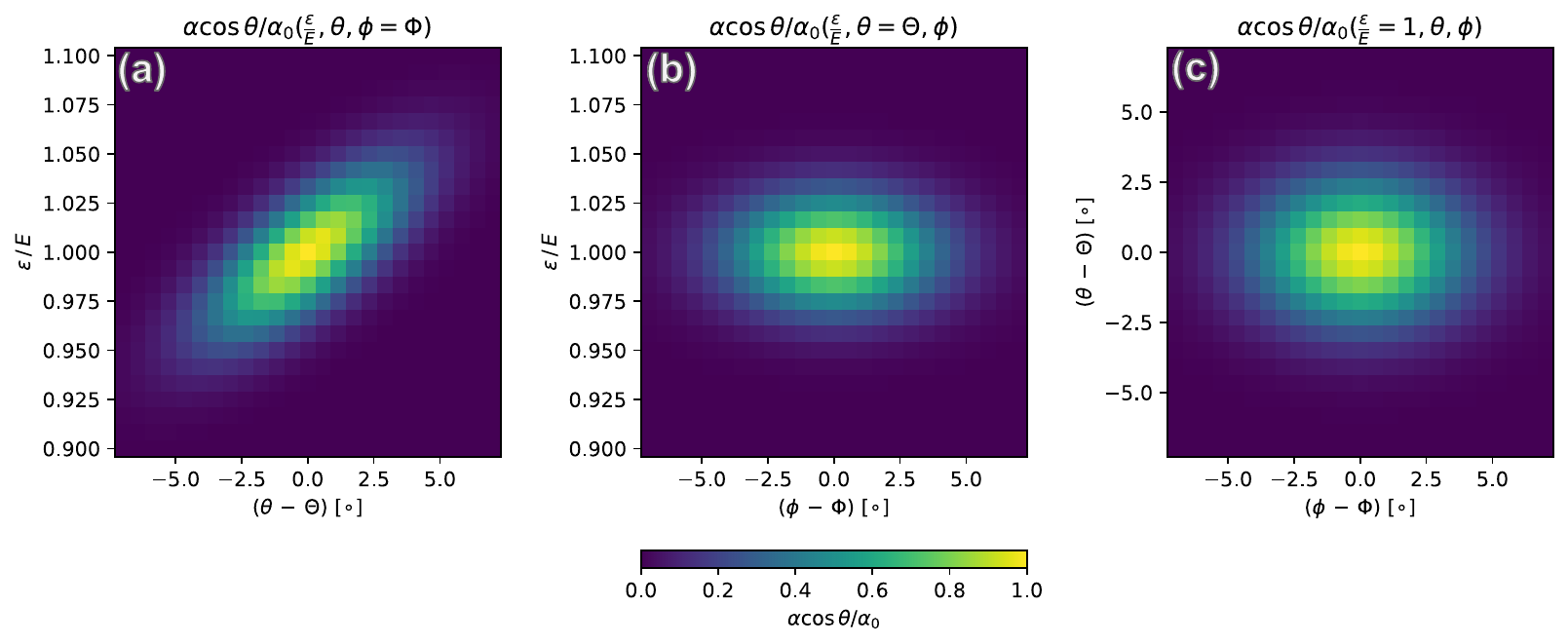}
    \caption{The $\alpha\cos{\theta}/\alpha_0$ of our instrument model as a function of \textbf{(a)} $\frac{\varepsilon}{E}$ and $\theta$, for $\phi=\Phi$, \textbf{(b)} $\frac{\varepsilon}{E}$ and $\phi$, for $\theta=\Theta$, and \textbf{(c)} $\theta$ and $\phi$, for $\frac{\varepsilon}{E}=1$.}
    \label{fig:efficiency}
\end{figure}

In each acquisition, the instrument records the number of particles in discrete $E$, $\Theta$, and $\Phi$ bins. The expected number of counts (recorded number of particles) in each $E$, $\Theta$, $\Phi$ bin, for a single acquisition is \citep{Lavraud2016,Nicolaou2019Turbulent}:

\begin{equation}
C_{\mathrm{exp}}(E,\Theta,\Phi)=\Delta\tau \int\limits_{\varepsilon_{\mathrm{min}}}^{\varepsilon_{\mathrm{max}}} \int\limits_{\theta_{\mathrm{min}}}^{\theta_{\mathrm{max}}} \int\limits_{\phi_{\mathrm{min}}}^{\phi_{\mathrm{max}}} \alpha(E,\Theta,\Phi,\varepsilon,\theta,\phi)f(\varepsilon,\theta,\phi)\frac{2}{m^2}\varepsilon\,\mathrm{d}\varepsilon\mathrm{cos}\theta\,\mathrm{d}\theta\,\mathrm{d}\phi ,
\label{eq:Counts}
\end{equation}
\noindent where $\Delta\tau$ is the duration of each acquisition and $\alpha(E,\Theta,\Phi,\varepsilon,\theta,\phi)$ is the effective aperture area which, in general, varies with the sampled energy and direction. The limits of the integral are determined by the minimum and maximum energy, elevation, and azimuth angle of the particles that can be detected in each bin.We now assume that for our concept instrument

\begin{equation}
\alpha(E,\Theta,\Phi,\varepsilon,\theta,\phi)\mathrm{cos}\theta=\alpha_{0}\exp\left[-\frac{\left(\frac{\varepsilon}{E}-1+\frac{\theta-\Theta}{S_{\mathrm{E\Theta}}}\right)^2}{2\left( \frac{\sigma_{E}}{E} \right)^2}\right]\exp\left[-\frac{\left( \theta - \Theta \right)^2}{2\left( \sigma_{\Theta} \right)^2}\right]\exp\left[-\frac{\left( \phi - \Phi \right)^2}{2\left( \sigma_{\Phi} \right)^2}\right],
\label{eq:response}
\end{equation}


\noindent where we consider the same $\alpha_{0}$ for each $E,\Theta,\Phi$ bin. For this study, we adjust $\alpha_0$, such that the peak of $C_{\mathrm{exp}}(E,\Theta,\Phi)$ is 10000 counts for each sample we simulate. The standard deviations $\sigma_{E}$, $\sigma_{\Theta}$, and $\sigma_{\Phi}$ describe the width of the transmission curves along $\varepsilon$, $\theta$, and $\phi$, respectively, within each $E,\Theta,\Phi$ bin. Our concept instrument has $\sigma_{E}\sim0.02E$, $\sigma_{\Theta}\sim2.55^{\circ}$, and $\sigma_{\Phi}\sim2.72^{\circ}$. Equation \ref{eq:response}, implies that the energy of the peak of the transmission depends on the elevation angle, which is a standard feature of electrostatic analyzers.\citep{Young2004,McComas2008SWAP,McComas2017}. This energy-elevation coupling of the response is adjusted by the $S_{\mathrm{E\Theta}}$ term, which in our model is set to $S_{\mathrm{E\Theta}}=120$, which simulates a response that is similar to the electron plasma spectrometer (CAPS/ELS) on Cassini \citep{Young2004,Nicolaou2022_Frontiers} and the Solar Wind Around Pluto (SWAP) on New Horizons \citep{McComas2008SWAP,Nicolaou2014b}. Fig. \ref{fig:efficiency}\textbf{(a)} shows $\alpha\cos{\theta}/\alpha_0$ of our model instrument as a function of $\frac{\varepsilon}{E}$ and $\theta$, for $\phi=\Phi$.  Fig. \ref{fig:efficiency}\textbf{(b)} shows $\alpha\cos{\theta}/\alpha_0$ as a function of $\frac{\varepsilon}{E}$ and $\phi$, for $\theta=\Theta$, and  Fig. \ref{fig:efficiency}\textbf{(c)} shows $\alpha\cos{\theta}/\alpha_0$ as a function of $\theta$ and $\phi$, for $\frac{\varepsilon}{E}=1$.

We simulate the expected number of counts in each $E,\Theta,\Phi$ bin, based on Eq. \ref{eq:Counts} and using the expressions for the effective aperture and response function as explained above. To solve the triple integral numerically, we substitute the integrals with sums, i.e.:

\begin{eqnarray}
C_{\mathrm{exp}}(E,\Theta,\Phi)&=&\Delta\tau \sum_{i=1}^{25} \sum_{j=1}^{25} \sum_{k=1}^{25}\alpha_{0}\exp\left[-\frac{\left(\frac{\varepsilon_{i}}{E}-1+\frac{\theta_{j}-\Theta}{S_{E\Theta}}\right)^2}{2\left( \frac{\sigma_{E}}{E} \right)^2}\right]\exp\left[-\frac{\left( \theta_{j} - \Theta \right)^2}{2\left( \sigma_{\Theta} \right)^2}\right]\exp\left[-\frac{\left( \phi_{k} - \Phi \right)^2}{2\left( \sigma_{\Phi} \right)^2}\right]\nonumber\\&\times&f(\varepsilon_{i},\theta_{j},\phi_{k})\frac{2}{m^2}\varepsilon_{i}\,\mathrm{d}\varepsilon_{i}\,\mathrm{d}\theta_{j}\,\mathrm{d}\phi_{k} ,
\label{eq:Counts_sums}
\end{eqnarray}
\noindent where we divide the acceptance width of each bin in discrete steps $\varepsilon_{i},\theta_{j},\phi_{k}$. Appendix \ref{app:optimization} shows how we optimize our model and decide to use 25$\times$25$\times$25 of $\varepsilon_{i}\times\theta_{j}\times\phi_{k}$ steps. In each bin, we assign a measurement $C(E,\Theta,\Phi)$, which is taken randomly from the Poisson distribution with expectation value $C_{\mathrm{exp}}(E,\Theta,\Phi)$. This is done to model the statistical uncertainty of each measurement.\citep{Nicolaou2020spaceweather,Nicolaou2020b,Nicolaou2023,Nicolaou2024apj,Nicolaou2024rasti} Nevertheless, the adjustment of $\alpha_0$ as explained above, reduces biases caused by statistical errors.

\subsection{Plasma distributions constructed from observations}
\label{sec:approximations}
To construct the 3D VDFs from in-situ observations, we treat the particle energy $\varepsilon$, elevation $\theta$, and azimuth $\phi$ as constants over the acceptance width of each bin and equal to their central values $E$, $\Theta$, and $\Phi$, respectively. Thus, the distribution function is $f(E,\Theta,\Phi)$, and also constant within the acceptance width of each $E,\Theta,\Phi$ bin. With this approximation, Eq. \ref{eq:Counts} becomes:

\begin{equation}
C_{\mathrm{exp}}(E,\Theta,\Phi) \sim f(E,\Theta,\Phi)\frac{2}{m^2}\Delta\tau E^2 \int\limits_{\varepsilon_{\mathrm{min}}}^{\varepsilon_{\mathrm{max}}} \int\limits_{\theta_{\mathrm{min}}}^{\theta_{\mathrm{max}}} \int\limits_{\phi_{\mathrm{min}}}^{\phi_{\mathrm{max}}}\alpha(E,\Theta,\Phi,\varepsilon,\theta,\phi)\,\frac{\mathrm{d}\varepsilon}{E}\mathrm{cos}\theta\,\mathrm{d}\theta\,\mathrm{d}\phi,
\label{eq:simple_Counts}
\end{equation}

\noindent where the integral on the right-hand term is the energy dependent, effective geometric factor of the instrument

\begin{equation}
G(E,\Theta,\Phi)\equiv\int\limits_{\varepsilon_{\mathrm{min}}}^{\varepsilon_{\mathrm{max}}} \int\limits_{\theta_{\mathrm{min}}}^{\theta_{\mathrm{max}}} \int\limits_{\phi_{\mathrm{min}}}^{\phi_{\mathrm{max}}}\alpha(E,\Theta,\Phi,\varepsilon,\theta,\phi)\,\frac{\mathrm{d}\varepsilon}{E}\mathrm{cos}\theta\,\mathrm{d}\theta\,\mathrm{d}\phi.
\label{eq:Gfactor_integral}
\end{equation}

\noindent Under this  simplification then,\citep{Lewis2008,Nicolaou2020spaceweather,Nicolaou2020b} the expected number of counts in each bin is\begin{equation}
C_{\mathrm{exp}}(E,\Theta,\Phi)\sim\frac{2}{m^2}G(E,\Theta,\Phi)E^2 f(E,\Theta,\Phi)\Delta \tau.
\label{eq:Cexp_simple}
\end{equation}
Assuming further that the obtained measurements $C(E,\Theta,\Phi)$  are representative of the expected counts, then it is straightforward to convert the observations to plasma distribution functions using:

\begin{equation}
f_{\mathrm{out}}(E,\Theta,\Phi)\sim\frac{m^2}{2G(E,\Theta,\Phi)E^2\Delta\tau}C(E,\Theta,\Phi).
\label{eq:fout_simple}
\end{equation}
Eq. \ref{eq:fout_simple}  fails to describe plasma measurements when the underlying distribution functions change significantly over the acceptance width of each bin of the instrument. In this study we investigate the accuracy of the approach used in Eq. \ref{eq:fout_simple}. In order to do that, we simulate observations $C(E,\Theta,\Phi)$ using a high-resolution model as described in Sec. \ref{sec:forward_model} and in Appendix \ref{app:optimization}, for Maxwellian proton distribution functions for a range of input bulk speeds $V_{\mathrm{in}}$, and temperatures $T_{\mathrm{in}}$. We then compare the differences between the constructed $f_{\mathrm{out}}(E,\Theta,\Phi)$ and the input distributions $f(E,\Theta,\Phi)$ and the differences between their velocity moments, as we explain in Sec. \ref{sec:residuals} below.

\subsection{Quantifying the inaccuracies}
\label{sec:residuals}
Our evaluation is based on the comparison between the input distribution functions $f(E,\Theta,\Phi)$ and the corresponding distributions we construct from the simulated observations $f_{\mathrm{out}}(E,\Theta,\Phi)$. For different sets of input parameters, we calculate the distribution of the residuals:
\begin{equation}
    F_{\mathrm{residual}}(E,\Theta,\Phi)=\mathrm{log}_{10}[f_{\mathrm{out}}(E,\Theta,\Phi)]-\mathrm{log}_{10}[f(E,\Theta,\Phi)],
\end{equation}
considering only $E,\Theta,\Phi$ bins with $C(E,\Theta,\Phi)\,>\,1$.
For each combination of the input plasma parameters, we calculate the mean absolute value of the residuals as
\begin{equation}
    R=\frac{1}{N_{\mathrm{E}}\times N_{\mathrm{\Theta}}\times N_{\mathrm{\Phi}}}\sum_{i=1}^{N_{\mathrm{E}}}\sum_{j=1}^{N_{\Theta}}\sum_{k=1}^{N_{\mathrm{\Phi}}}|F_{\mathrm{residual}}(E_{i},\Theta_{j},\Phi_{k})|,
\label{eq:R}
\end{equation}
where indices $i,j,k$ now indicate the individual energy, elevation, and azimuth bins of the instrument with $C(E,\Theta,\Phi)\,>\,1$. Finally, in order to estimate the impact of the limited instrument resolution to the plasma bulk parameters, for each set of input plasma parameters, we compare the velocity moments of $f_{\mathrm{out}}(E,\Theta,\Phi)$ and the corresponding velocity moments of $f(E,\Theta,\Phi)$. We calculate the first three orders of velocity moments of each $f_{\mathrm{out}}$ and $f$, determining the corresponding densities $N_{\mathrm{out}}$ and $N_{\mathrm{f}}$, speeds $V_{\mathrm{out}}$ and $V_{\mathrm{f}}$, and temperatures $T_{\mathrm{out}}$ and $T_{\mathrm{f}}$ (see Appendix \ref{app:moments}). Although $f(E,\Theta,\Phi)$ is the value of the input distribution at the center of each $E,\Theta,\Phi$ bin, we do not expect the determined moments $N_{\mathrm{f}}$, $V_{\mathrm{f}}$, and $T_{\mathrm{f}}$ to be identical to their corresponding input parameters $N_{\mathrm{in}}$, $V_{\mathrm{in}}$, and $T_{\mathrm{in}}$, due to the limited sampling of the distribution.\citep{Nicolaou2019Turbulent,Nicolaou2020spaceweather}. However, we expect that $N_{\mathrm{f}}$, $V_{\mathrm{f}}$, and $T_{\mathrm{f}}$ would be identical to the corresponding moments of $f_{\mathrm{out}}$ for cases with negligible error. Thus, for the purposes of this study we investigate the ratios $\frac{N_{\mathrm{out}}}{N_{\mathrm{f}}}$, $\frac{V_{\mathrm{out}}}{V_{\mathrm{f}}}$, and $\frac{T_{\mathrm{out}}}{T_{\mathrm{f}}}$.

\section{Results}
\label{sec:results}
\begin{figure}
    \centering
    \includegraphics[width=1\linewidth]{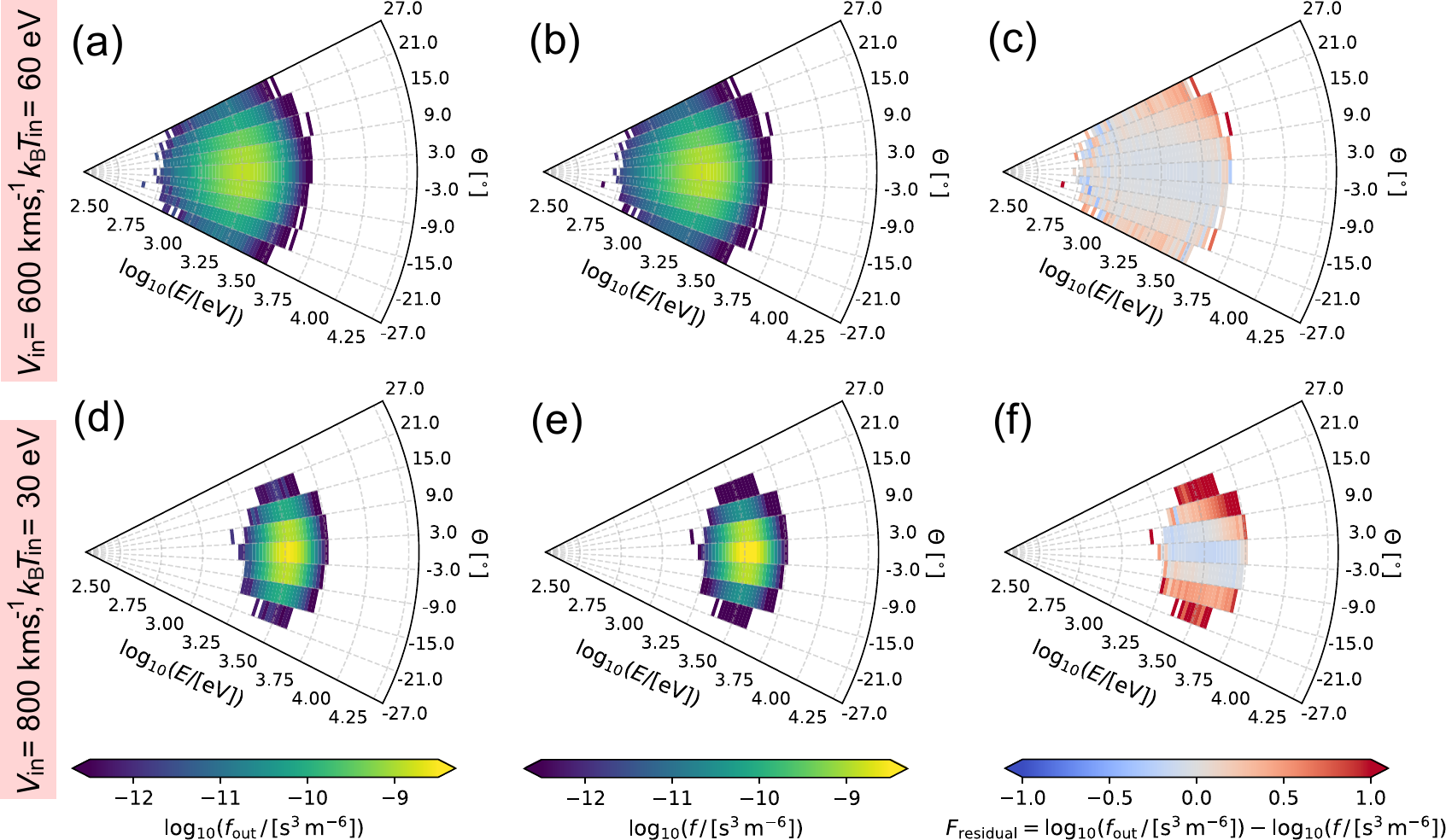}
    \caption{2D-cuts of a \textbf{(a)} constructed $f_{\mathrm{out}}(E,\Theta,\Phi=0^{\circ})$, \textbf{(b)} input $f(E,\Theta,\Phi=0^{\circ})$, and \textbf{(c)} the residual $F_{\mathrm{residual}}(E,\Theta,\Phi=0^{\circ})$ distributions, for simulated plasma with $N_{\mathrm{in}}=10\mathrm{\,cm}^{-3}$, $V_{\mathrm{in}}$ = 600 km$\,\mathrm{s^{-1}}$ and $k_{\mathrm{B}}T_{\mathrm{in}}$ = 60 eV. \textbf{(d)}, \textbf{(e)}, and \textbf{(f)} are the corresponding distributions for plasma with the same density but $V_{\mathrm{in}}$ = 800 km$\,\mathrm{s^{-1}}$ and $k_{\mathrm{B}}T_{\mathrm{in}}$ = 30 eV.}
    \label{fig:residual_examples}
\end{figure}

Figure \ref{fig:residual_examples}\textbf{a} shows a 2D cut of one $f_{\mathrm{out}}(E,\Theta,\Phi=0^{\circ})$, constructed from simulated observations of plasma protons with $N_{\mathrm{in}}=10 \mathrm{\,cm}^{-3}$, $V_{\mathrm{in}}=600\mathrm{\,km\,s}^{-1}$ and $k_{\mathrm{B}}T_{\mathrm{in}}=60$ eV. Figure \ref{fig:residual_examples}\textbf{b} shows the input distribution $f(E,\Theta,\Phi=0^{\circ})$ for the same plasma parameters and \ref{fig:residual_examples}\textbf{c} shows the residual distribution $F_{\mathrm{residual}}(E,\Theta,\Phi=0^{\circ})$. For this set of input plasma parameters, the distribution extends beyond the elevation field of view. At first glance, $f_{\mathrm{out}}(E,\Theta,\Phi=0^{\circ})$ and $f(E,\Theta,\Phi=0^{\circ})$ appear very similar. However, $F_{\mathrm{residual}}(E,\Theta,\Phi=0^{\circ})$ is negative at the core (at velocities near the peak of $f$ and $f_{\mathrm{out}}$) and positive at the tails (velocities away from the peak). This is implying that the peak of $f_{\mathrm{out}}(E,\Theta,\Phi=0^{\circ})$ is less than the peak of $f(E,\Theta,\Phi=0^{\circ})$. On the other hand, $f_{\mathrm{out}}(E,\Theta,\Phi=0^{\circ})$ is greater than $f(E,\Theta,\Phi=0^{\circ})$ at the tails.  
Panels \textbf{d}, \textbf{e}, and \textbf{f} show $f_{\mathrm{out}}(E,\Theta,\Phi=0^{\circ})$, $f(E,\Theta,\Phi=0^{\circ})$, and their residuals $F_{\mathrm{residual}}(E,\Theta,\Phi=0^{\circ})$, for protons with the same density, but for $V_{\mathrm{in}}=800\mathrm{\,km\,s}^{-1}$ and $k_{\mathrm{B}}T_{\mathrm{in}}=30$ eV.  For this set of input parameters, the distribution function does not extend beyond the instrument's field of view. Similarly to the previous example, $f_{\mathrm{out}}$ underestimates the core and overestimates the tails of the input distribution. In this case, we can directly observe differences between $f_{\mathrm{out}}$ and $f$, by comparing panels \textbf{d} and \textbf{e}. Moreover, $F_{\mathrm{residual}}(E,\Theta,\Phi=0^{\circ})$ in \ref{fig:residual_examples}\textbf{f} extends to bigger absolute values than the corresponding $F_{\mathrm{residual}}(E,\Theta,\Phi=0^{\circ})$ of the slower and hotter plasma example shown in \ref{fig:residual_examples}\textbf{c}.

\begin{figure}
    \centering
    \includegraphics[width=0.5\linewidth]{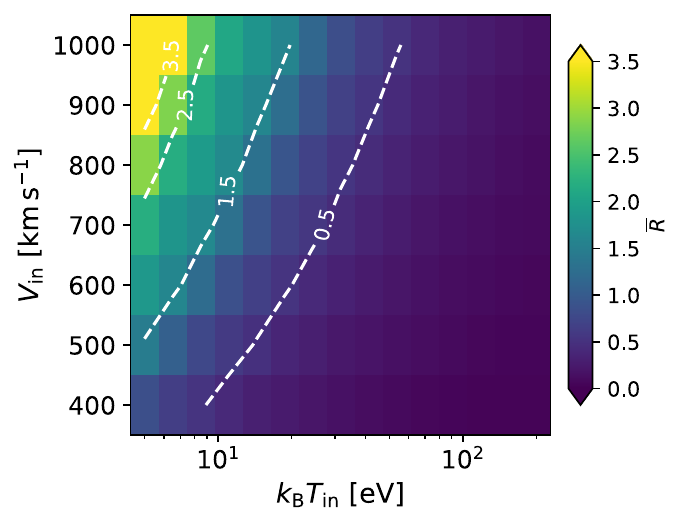}
    \caption{2D histogram of the averaged residuals $\overline{R}$ as functions of the input speed $V_{\mathrm{in}}$ and temperature $k_{\mathrm{B}}T_{\mathrm{in}}$. The white dashed lines are contours of selected $\overline{R}$ values (see text for details).}
    \label{fig:residuals}
\end{figure}

We complete our evaluations by calculating the mean residuals $R$ (Eq. \ref{eq:R}), for a wide range of input solar wind proton bulk speeds $V_\mathrm{in}$ and temperatures $T_{\mathrm{in}}$, typical for protons in the inner heliosphere.\citep{Freeman1988} For all simulations, we use $N_{\mathrm{in}}=10\mathrm{\,cm}^{-3}$. For each set of input plasma parameters, we simulate 10 samples. Thus, for each $V_{\mathrm{in}}$ - $T_{\mathrm{in}}$ set, we calculate ten $R$ values and eventually, their average $\overline{R}$ (average over the ten samples). Figure \ref{fig:residuals}
shows $\overline{R}$ as a function of $V_{\mathrm{in}}$ and $T_{\mathrm{in}}$. The white curves are contours of selected $\overline{R}$ values. We observe that $\overline{R}$ increases with increasing speed and/or decreasing temperature. For the fastest ($V_{\mathrm{in}}$=1000 km$\,\mathrm{s}^{-1}$) and coldest ($k_{\mathrm{B}}T_{\mathrm{in}}$ = 5 eV)  distribution we examine here, $\overline{R}$ is greater than 3.5. This means that for this set of input parameters, the difference between the constructed and input distributions is several orders of magnitude, on average. Even for a relatively slow solar wind with $V_{\mathrm{in}}$ = 400 km$\,\mathrm{s}^{-1}$, we see a rather  significant difference ($\overline{R}$ > 0.5) for temperatures $k_{\mathrm{B}}T_{\mathrm{in}} < 9$ eV.     

\begin{figure}
    \centering
    \includegraphics[width=1\linewidth]{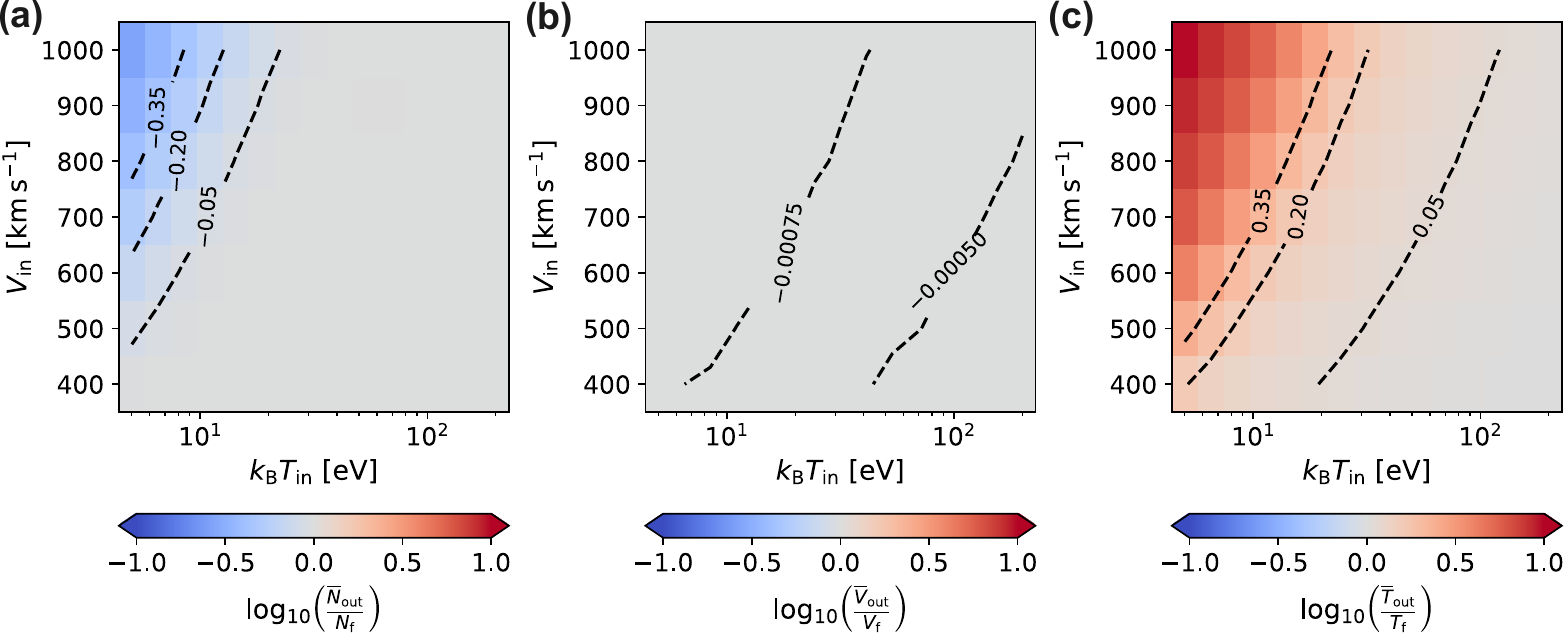}
    \caption{Ratios of the plasma parameters determined from the constructed distributions, over the corresponding parameters underlying the input distributions; 2D histograms of \textbf{(a)} $\mathrm{log_{10}}\left( \frac{\overline{N}_{\mathrm{out}}}{N_{\mathrm{f}}} \right)$, \textbf{(b)} $\mathrm{log_{10}}\left( \frac{\overline{V}_{\mathrm{out}}}{V_{\mathrm{f}}} \right)$, and \textbf{(c)} $\mathrm{log_{10}}\left( \frac{\overline{T}_{\mathrm{out}}}{T_{\mathrm{f}}} \right)$ as functions of $V_{\mathrm{in}}$ and $T_{\mathrm{in}}$. The black dashed curves are contours of selected ratio values.}
    \label{fig:moments_maps}
\end{figure}

In Figure \ref{fig:moments_maps}\textbf{a-c}, we show 2D histograms of the average output density, speed and temperature (average of the values determined for each of the 10 simulated samples per $V_{\mathrm{in}}-T_{\mathrm{in}}$ set), divided by the corresponding moment of the input distribution, for each set of input parameters. In all panels, the ratios are $\sim$1 for the smallest bulk speed and the largest plasma temperature we examine here, which are $V_{\mathrm{in}}$= 400 kms\textsuperscript{-1} and $T_{\mathrm{in}}$= 200 eV, respectively. According to Figure \ref{fig:moments_maps}\textbf{(a)}, as the proton speed increases and/or the plasma proton temperature decreases, the constructed distribution integrates to a smaller density than the one underlying the input distribution. There are examples within the examined range of parameters, for which the density determined by $f_{\mathrm{out}}$ is underestimated by more than 50\% ($\mathrm{log_{10}}\left(\frac{\overline{N}_{\mathrm{out}}}{N_{\mathrm{f}}}\right) <$ -0.3). According to Figure \ref{fig:moments_maps}\textbf{(b)}, there is negligible difference between the speed determined by $f_{\mathrm{out}}$ and the speed underlying $f$. For all $V_{\mathrm{in}}-T_{\mathrm{in}}$ we examine here, the difference is much less than 1\% ($\mathrm{log_{10}}\left(\frac{\overline{V}_{\mathrm{out}}}{V_{\mathrm{f}}}\right) <$ -0.001). According to Figure \ref{fig:moments_maps}\textbf{(c)}, the temperatures determined by $f_{\mathrm{out}}$ are significantly overestimated for a wide range of input speeds and temperatures. For instance, even for the slowest solar wind case ($V_{\mathrm{in}} \sim 400\mathrm{km\,s^{-1}}$) the temperature ratio is greater than 1.12 ($\mathrm{log_{10}}\left(\frac{T_{\mathrm{out}}}{T_{\mathrm{f}}}\right) >$ 0.05) for all input temperatures below 20 eV. For the fastest and coldest solar wind example we simulate here, the temperature is overestimated by a factor of ten.

\section{Discussion}
\label{sec:discussion}
Our results show that the interpretation of plasma observations by electrostatic analyzers may suffer significant inaccuracies, caused by the incapability of instruments to resolve the shape of the plasma VDFs within the instrument's energy, and/or angular bins. We model single-species plasma observations by an electrostatic analyzer concept and demonstrate that these systematic errors are larger as the bulk speed increases and/or the temperature decreases (see Figure \ref{fig:residuals}). 

For the same plasma conditions, observations by analyzers with lower resolution will return VDFs with larger uncertainties. When we refer to the instrument resolution in this study, we refer to the widths of the transmission curves along $\varepsilon$, $\theta$, and $\phi$, which are given by $\sigma_{E}$, $\sigma_{\Theta}$, and $\sigma_{\Phi}$, respectively (see Section \ref{sec:forward_model}). Therefore, for each instrument with specific energy and angular acceptance widths, there is a certain range of plasma parameters for which the constructed VDFs are reliable. We argue that in order to guarantee the validity of science studies, it is important to estimate the confidence level of the VDFs and their products that are determined from ESA observations. Such evaluation is possible by applying the same methodology we present here, to specific ESAs and plasma distribution functions.

\subsection{VDF shape and instrument resolution }
\begin{figure}
    \centering
    \includegraphics[width=0.9\linewidth]{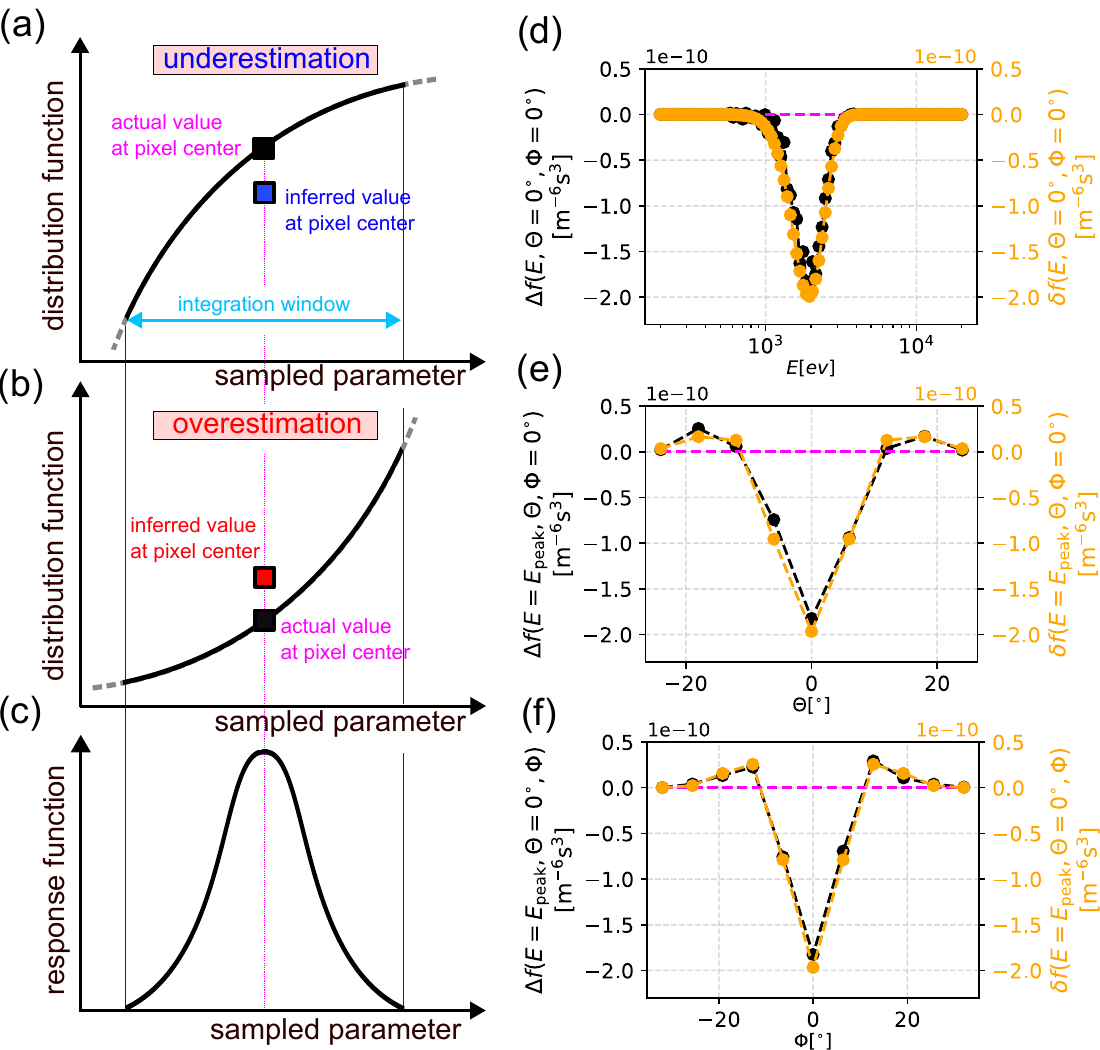}
    \caption{\textbf{(a)} An example of a distribution function that has a negative second-order derivative with respect to the sampled parameter within the bin, which results in an underestimation of the distribution, and \textbf{(b)} an example of a distribution with positive second-order derivative with respect to the sampled parameter within the bin, which results in an overestimation of the plasma distribution.  \textbf{(c)} A symmetric, Gaussian response function of a bin along the sampled parameter. \textbf{(d)} 1D cuts of the difference between the constructed and input plasma VDFs, $\Delta f$ (black), and the analytical expression $\delta f$ (orange), as functions of energy and for the elevation and azimuth of the peak, considering plasma with $N_{\mathrm{in}}=10\mathrm{\,cm}^{-3}$, $V_{\mathrm{in}}$ = 600 km$\,\mathrm{s^{-1}}$, and $k_{\mathrm{B}}T_{\mathrm{in}}$ = 60 eV. \textbf{(e)} 1D cuts of $\Delta f$ and $\delta f$, as functions of the elevation flow direction, at the energy and azimuth of the peak, and \textbf{(f)} 1D cuts of $\Delta f$ and $\delta f$, as functions of the azimuth flow direction, at the energy and elevation direction of the peak, for the same plasma conditions.}
    \label{fig:derivatives_sketch}
\end{figure}

In Figure \ref{fig:derivatives_sketch}, we demonstrate how the unresolved shape of the VDFs within the instrument's bins causes the systematic uncertainties we examine in this study. Panels \textbf{(a)} and \textbf{(b)} show two examples of an input distribution function shape along one of the sampled parameters (either energy or angle) within a single bin. Panel \textbf{(c)} shows the Gaussian response as a function of the sampled parameter within the bin. In the example shown in panel \ref{fig:derivatives_sketch}\textbf{a}, the input distribution function increases as the sampled parameter increases. However, the positive gradient of the distribution decreases (negative second-order derivative). This results in an asymmetric distribution with respect to its value at the center of the bin. The bigger contribution to the flux integral (Eq. \ref{eq:Counts}) comes from $f$ values that are smaller than the value of $f$ at the center of the bin. As a consequence, the observed number of counts is smaller than the counts according to Eq. \ref{eq:Cexp_simple} using the value of $f$ at the center of the bin. Therefore, the $f_{\mathrm{out}}$ constructed with Eq. \ref{eq:fout_simple} underestimates the actual distribution $f$ at the bin center. 

The case shown in \ref{fig:derivatives_sketch}\textbf{(b)} has a positive second-order derivative. In this case, the asymmetry of the distribution results in a larger number of counts compared to those Eq. \ref{eq:Cexp_simple} estimates with the value of $f$ at the bin center. As a result, Eq. \ref{eq:fout_simple} overestimates the VDF. 

In Appendix \ref{app:derivatives}, we derive the Taylor series of an isotropic Maxwellian VDF $f(\varepsilon,\theta,\phi)$, up to second-order terms, and evaluate it at the instrument bin centers $E,\Theta,\Phi$. We demonstrate that up to second-order terms the systematic differences between $f_{\mathrm{out}}$ and $f$ are approximately 
\begin{equation}
\delta f(E,\Theta,\Phi)\approx\frac{1}{2}\left[ \sigma_{E}^2\frac{\partial^2f}{\partial \varepsilon^2}(E,\Theta,\Phi)+\sigma_{\Theta}^2\frac{\partial^2f}{\partial\theta^2}(E,\Theta,\Phi)+\sigma_{\Phi}^2\frac{\partial^2f}{\partial\phi^2}(E,\Theta,\Phi) \right],
\label{eq:deltaf_inst}
\end{equation}
which implies that indeed, the systematic uncertainties increase with increasing second-order derivatives of $f$. Eq. \ref{eq:deltaf_inst} shows that for the same $f$, the uncertainties increase with increasing $\sigma_{E}$, $\sigma_{\Theta}$ and $\sigma_{\Phi}$, which determine the instrument's energy and angular resolution. Appendix \ref{app:derivatives} shows the derivation of Eq. \ref{eq:deltaf_inst} and the analytical expressions for the derivatives of $f$. 

In Figure \ref{fig:derivatives_sketch}\textbf{d-f}, we compare 1D curves of the analytical $\delta f$ function with the corresponding 1D cuts of 
\begin{equation}
\Delta f(E,\Theta,\Phi)=f_{\mathrm{out}}(E,\Theta,\Phi)-f(E,\Theta,\Phi).
\end{equation}
\noindent The black curve in \ref{fig:derivatives_sketch}\textbf{(d)} is an 1D cut of $\Delta f$ at the elevation and azimuth bins for which the distribution has its peak, i.e. $\Delta f(E,\Theta=0^{\circ}$,$\Phi=0^{\circ})$, and considering plasma with $N_{\mathrm{in}}=10\mathrm{\,cm}^{-3}$, $V_{\mathrm{in}}$ = 600 km$\,\mathrm{s^{-1}}$ and $k_{\mathrm{B}}T_{\mathrm{in}}$ = 60 eV. The orange curve in the same panel shows the 1D cut $\delta f(E,\Theta=0^{\circ},\Phi=0^{\circ})$, calculated analytically for the same plasma conditions. In Fig. \ref{fig:derivatives_sketch}
\textbf{e}, we show $\Delta f(E=E_{\mathrm{peak}},\Theta,\Phi=0^{\circ})$ and $\delta f(E=E_{\mathrm{peak}},\Theta,\Phi=0^{\circ})$ for the same plasma conditions, which are the 1D cuts of $\Delta f$ and $\delta f$, at the energy and azimuth bins which capture the peak of $f$, respectively. Fig. \ref{fig:derivatives_sketch}
\textbf{f}, shows the corresponding 1D-cuts at $E=E_{\mathrm{peak}}$ and $\Theta=0^{\circ}$.  The apparent similarity between $\Delta f$ and $\delta f$ confirms that Eq. \ref{eq:deltaf_inst} estimates successfully the uncertainties in this example. As the higher-order derivatives of the VDF increase (colder and/or faster species), and as the instrument resolution decreases (larger $\sigma_{E}$, $\sigma_{\Theta}$, $\sigma_{\Phi}$ increase), Eq. \ref{eq:deltaf_inst} would require higher-order terms to describe the uncertainties. In Appendix \ref{app:derivatives}, we explain the approach behind the derivation of Eq. \ref{eq:deltaf_inst}, which is useful for fast and easy diagnosis of the level of expected uncertainties.  

\subsection{Impact on plasma physical parameters}
Figure \ref{fig:residual_examples} shows that the constructed distributions, in general, underestimate the core of the input distribution functions, while they overestimate their tails. This is in agreement with the diagrams in Figure \ref{fig:derivatives_sketch}\textbf{d-f}. It is also consistent with the fact that in colder and faster plasmas, the plasma density is underestimated and the temperature is overestimated (Figure \ref{fig:moments_maps}). The core of the distribution contributes significantly to the zeroth order velocity moment (the particle density), while the tails, contain the higher energy particles which make a significant contribution to the second order velocity moment (the temperature of the species). The bulk speed accuracy is barely affected in the examples we examine here. Even in the colder and faster plasma examples we examine, and for the specific instrument resolution, the residuals are approximately symmetric around the bulk (see Figs. \ref{fig:residual_examples} and \ref{fig:derivatives_sketch}) and thus, the first-order velocity moment is barely affected. We do not expect this to hold for any type of $f$ or for bigger $\sigma_{E}$.  This study does not examine the accuracy of the recovered plasma parameters for different plasma bulk velocity directions. Given the typical Gaussian response of the individual elevation and azimuth bins, we expect different distribution of counts as a cold/fast proton beam shifts in direction (in sub-bin scales). 

We acknowledge that the systematic errors in the VDF shapes can have a vital impact on scientific studies which require detailed knowledge of plasma VDFs \citep[e.g.][]{Wilson2022}. Our results demonstrate that the VDF shape of cold and fast solar wind protons is highly affected, even by orders of magnitude. This systematic uncertainty is a function of the input plasma, and thus, it is expected to lead to erroneous correlations between the plasma parameters; i.e. artificially larger VDF tails in colder and faster wind.  There are cases within the range of plasma parameters  we examine, in which the systematic uncertainties of this type exceed significantly the statistical and systematic uncertainties of different sources, such as background noise \citep[e.g.][]{Nicolaou2022noise,Nicolaou2023}, plasma fluctuations \citep[e.g.][]{Nicolaou2019Turbulent}, count uncertainties \citep[e.g.][]{Nicolaou2020,Nicolaou2020c}, limited sampling \citep[][]{Nicolaou2020spaceweather},  and the incapability to distinct between VDFs of different species. \citep[e.g.][]{Zhang2024} 

We highlight that a critical evaluation of the uncertainties in specific applications should account for the VDFs of all the species that the instrument detects. For instance, ESAs in the solar wind and planetary magnetosheaths capture the distributions of alpha particles along with those of the protons. \citep[e.g.][]{Stansby2019,Bruno2024,DeWeese2022,Vech2021}. For co-moving proton and alpha populations, the VDFs of alphas extend at higher energy-per-charge bins than those recording the proton VDFs, due to their higher bulk energy-per-charge at the same velocities. Higher energy-per-charge bins however, have larger $\sigma_{E}$ and thus, even if the VDFs of the two species had the same shape, and even if the analysis could distinguish between the two species, the VDFs of alphas would be resolved with larger systematic uncertainty compared to protons.


\subsection{Potential mitigations}
One popular technique to determine the plasma VDFs, is by fitting the observations to forward model predictions. \citep[e.g.][]{Wilson2008,Wilson2015,Nicolaou2020spaceweather} With this technique, we can optimize the parameters of analytical VDF models to reproduce the actual observations. We argue that the use of high-resolution forward models which take into account the detailed response function of the instruments and the VDF shapes on sub-bin scales as we describe in Section \ref{sec:forward_model}, can overcome the systematic uncertainties arising from the instrument's finite resolution. Although this is one possible way to recover the actual VDFs, it requires a detailed implementation of the instrument's response function per bin and a numerical calculation on sub-bin scales. 

Solar wind protons usually exhibit non-thermal features, such as beams and supra-thermal tails \citep[e.g.][]{Marsch2012,Livadiotis2013,Verscharen_review}, and an accurate forward modeling would require numerous iterations with a variety of input $f$ functions, beyond the isotropic Maxwell distribution. The users of forward models should keep in mind that the optimization of VDF models that do not correspond to the actual plasma VDF, leads to systematic errors.\citep[][]{Nicolaou_Livadiotis2016} Additionally, classic fitting techniques that are used for optimizing models to observations may introduce biases and lead to systematic errors and artificial correlations between the determined plasma parameters. \citep[e.g.][]{Stoneking1997,Nicolaou2024rasti,Nicolaou2024apj}

The results of this study are linked to the specific instrument model and under the specific plasma conditions we consider for our demonstrations, which are described in Section \ref{sec:methodology}. Our purpose is to notify the community that the accurate determination of plasma parameters from in-situ observations requires a thorough examination of the possible VDFs and knowledge of the instrument response function and resolution. The same technique we describe here can be adapted to evaluate the performance of any analyzer of a similar design, in any plasma conditions.


%
%

%

\section*{Author Declarations}
Conflict of Interest: The authors have no conflicts of interest to disclose.

\section*{Author Contributions}
\textbf{Georgios Nicolaou:} Conceptualization (lead); Formal analysis (lead); Investigation (lead); Methodology (lead); Validation (lead); Visualization (lead); Writing - original draft (lead); Writing, review \& editing (equal). \textbf{Charalambos Ioannou:} Investigation (support); Methodology (support); Visualization (support); Writing, review \& editing (equal). \textbf{Christopher Owen:} Investigation (support); Methodology (support); Visualization (support); Writing, review \& editing (equal). \textbf{Daniel Verscharen:} Investigation (support); Methodology (support); Visualization (support); Writing, review \& editing (support). \textbf{Andrei Fedorov:} Investigation (support); Methodology (support). \textbf{Philip Louarn:} Investigation (support); Methodology (support) .

\begin{acknowledgments}
CI acknowledges support from STFC PhD Studentship ST/X508858/1. CJO and DV acknowledge support from STFC Consolidated Grant ST/W001004/1. This work was partially supported by the Royal Society (UK) and the Consiglio Nazionale delle Ricerche (Italy) through the International Exchanges Cost Share scheme/Joint Bilateral Agreement project “Multi-scale electrostatic energisation of plasmas: comparison of collective processes in laboratory and space” (award numbers IEC$\backslash{R2}\backslash{222050}$ and SAC.AD002.043.021).  
\end{acknowledgments}

\section*{Data Availability Statement}
This study uses only simulated data which will be provided by the corresponding author upon request.

\appendix
\section{Model optimization}
\label{app:optimization}
The accuracy of the forward model increases as we increase the number of discrete $\varepsilon_i$, $\theta_j$, and $\phi_k$ steps within the bin width, at which we evaluate the sum in Eq. \ref{eq:Counts_sums}. By increasing the number of steps, the computational time increases. Thus, we optimize the model by using the minimum number of steps required for accurate simulations. As shown in Fig. 1, colder and faster distributions require a model with higher resolution to maintain a high accuracy of the simulated counts. For our evaluation, we first use 33 steps for each parameter ($\varepsilon$, $\theta$, $\phi$) to simulate the number of counts for an input Maxwellian with $N_{\mathrm{in}}=10\mathrm{\,cm^{-3}}$, $V_{\mathrm{in}}$ = 1000 km$\,\mathrm{s}^{-1}$ along $\Theta=\Phi=0^{\circ}$, and $T_{\mathrm{in}}$ = 1 eV. We also set $\alpha_0=1 \mathrm{\,m^2}$, in order to have counts recorded by many instrument bins. We use this simulation product as the high-resolution reference model (M33 model product). We then simulate measurements with models of different resolution, starting from a low number of $\varepsilon_i$, $\theta_j$, $\phi_k$ steps (same number of steps for each parameter) and simulate the number of counts of the same distribution function. We compare the output of each model with the product of M33. We calculate the chi-squared value $\chi^2$, the Pearson correlation coefficient, and the slope between the number of counts by each model and the M33 reference model. 

\begin{figure}
    \centering
    \includegraphics[width=1\linewidth]{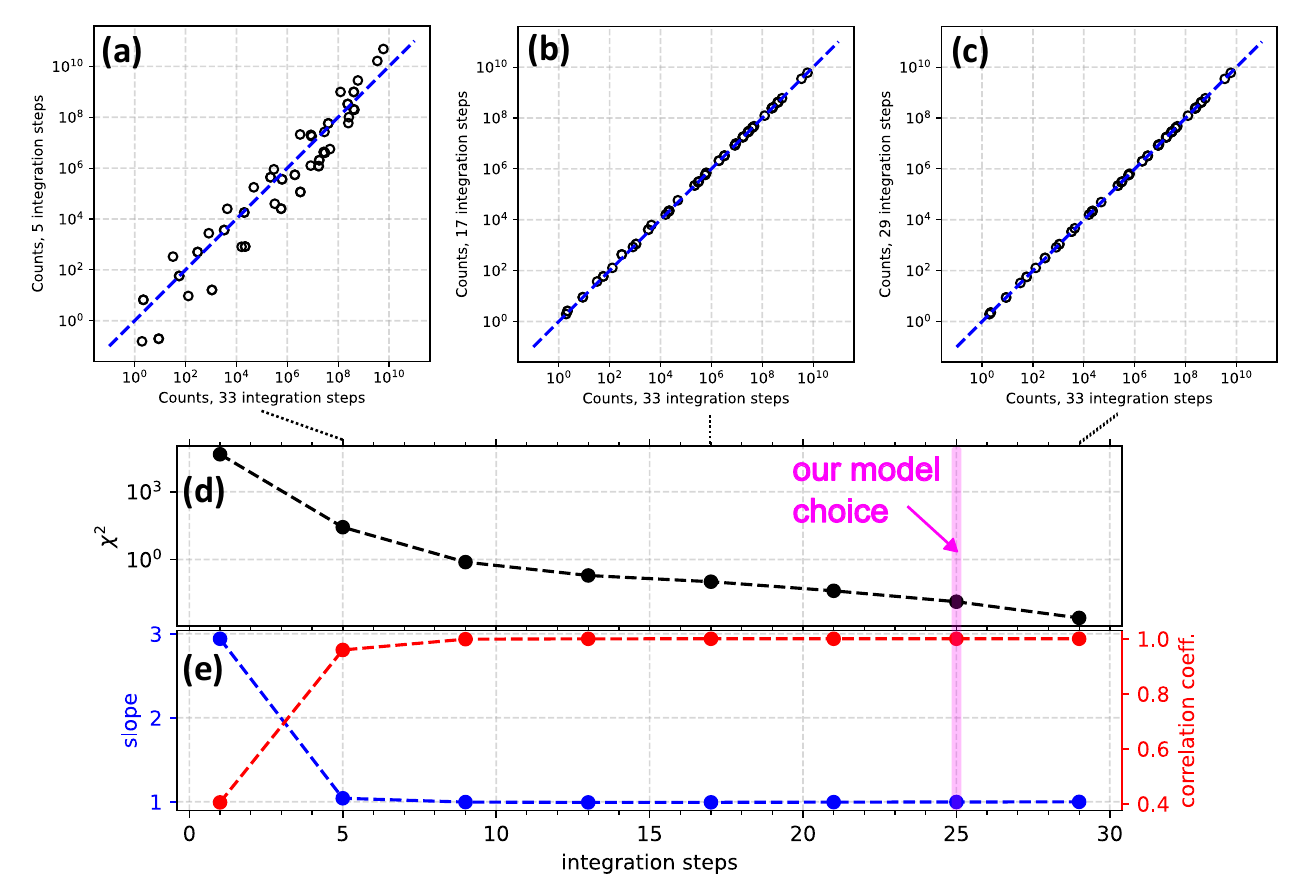}
    \caption{Comparison of observations produced by models of different resolution against the simulations of the M33 model. Number of counts simulated using \textbf{a}) 5 integration steps, \textbf{b}) 17, and \textbf{c}) 29 integration steps, versus the number of counts simulated by M33 for the same input proton plasma parameters. \textbf{d}) $\chi^2$ value of simulated counts by models of different integration steps and the counts simulated by M33 and \textbf{e}) the slope (blue) and the correlation coefficient (red) of the simulated counts by different models and M33, as functions of the integration steps of each model.}
    \label{fig:model_optimization}
\end{figure}

In Fig. \ref{fig:model_optimization}, we show the results of our model optimization. To optimize between computational time and accuracy, we use a model with 25 integration steps throughout this study, which produces virtually the same results as M33, for this fast and significantly cold Maxwellian we use for input. The model we use leads to $\chi^2\,\sim\,10^{-2}$, and it correlates almost perfectly with M33, since the Pearson correlation coefficient and the slope are both very close to 1. 

\section{Velocity moments}
\label{app:moments}

We calculate the output plasma bulk parameters as the velocity moments of the distribution function constructed from the observations.\citep{Nicolaou2020spaceweather,Zhang2024} The output plasma density is estimated by the $0^{th}$ order moment:
\begin{equation}
    N_{\mathrm{out}}=\sum_{i=1}^{96}\sum_{j=1}^{9}\sum_{k=1}^{11}\sqrt{2}f_{\mathrm{out}}(E_{i},\Theta_{j},\Phi_{k})\left(\frac{E_i}{m}\right)^{\frac{3}{2}}\mathrm{cos}\Theta_j\frac{\Delta E}{E} \Delta \Theta \Delta \Phi,
\end{equation}
where $\Delta E$, $\Delta \Theta$, and $\Delta \Phi$ are the differences between consecutive energy, elevation, and azimuth bin centers, respectively. The energy bins are exponentially spaced, resulting in a constant $\frac{\Delta E}{E}\approx0.05$, while the elevation and azimuth bins are uniformly spaced, such that $\Delta \Theta=6^{\circ}$ and $\Delta \Phi = 6.4^{\circ}$.
The first order moments determine the bulk velocity components of the plasma:
\begin{equation}
    V_{\mathrm{x,out}}=\frac{1}{N_{\mathrm{out}}}\sum_{i=1}^{96}\sum_{j=1}^{9}\sum_{k=1}^{11} 2 f_{\mathrm{out}}(E_{i},\Theta_{j},\Phi_{k}) \left(\frac{E_i}{m} \right)^2 \mathrm{cos}^2\Theta_j\mathrm{cos}\Phi_k\frac{\Delta E}{E} \Delta \Theta \Delta \Phi,
\end{equation}
\begin{equation}
    V_{\mathrm{y,out}}=\frac{1}{N_{\mathrm{out}}}\sum_{i=1}^{96}\sum_{j=1}^{9}\sum_{k=1}^{11}2f_{\mathrm{out}}(E_{i},\Theta_{j},\Phi_{k}) \left(\frac{E_i}{m} \right)^2 \mathrm{cos}^2\Theta_{j} \mathrm{sin}\Phi_k  \frac{\Delta E}{E} \Delta \Theta \Delta \Phi,
\end{equation}
and
\begin{equation}
    V_{\mathrm{z,out}}=\frac{1}{N_{\mathrm{out}}}\sum_{i=1}^{96}\sum_{j=1}^{9}\sum_{k=1}^{11}2f_{\mathrm{out}}(E_{i},\Theta_{j},\Phi_{k}) \left(\frac{E_i}{m}\right)^2 \mathrm{cos}\Theta_j\mathrm{sin}\Theta_{j}\frac{\Delta E}{E} \Delta \Theta \Delta \Phi,
\end{equation}
from which we obtain the bulk speed:
\begin{equation}
V_{\mathrm{out}}=\sqrt{  V_{\mathrm{x,out}}^2+  V_{\mathrm{y,out}}^2+  V_{\mathrm{z,out}}^2}.
\end{equation}
The second order moment determines the scalar temperature:
\begin{equation}
k_{\mathrm{B}}T_{\mathrm{out}}=\frac{1}{3N_{\mathrm{out}}}\sum_{i=1}^{96}\sum_{j=1}^{9}\sum_{k=1}^{11}\sqrt{\frac{2}{m}} \left(w_{\mathrm{x},ijk}^2+w_{\mathrm{y},ijk}^2+w_{\mathrm{z},ijk}^2 \right)f_{\mathrm{out}}(E_{i},\Theta_{j},\Phi_{k})E_i^{\frac{3}{2}}\mathrm{cos}\Theta_j \frac{\Delta E}{E} \Delta \Theta \Delta \Phi,
\end{equation}

\noindent where
\begin{equation}
w_{\mathrm{x},ijk}=\sqrt{\frac{2E_i}{m}}\mathrm{cos}\Theta_j\mathrm{cos}\Phi_k-V_{\mathrm{x,out}},
\end{equation}
\begin{equation}
w_{\mathrm{y},ijk}=\sqrt{\frac{2E_i}{m}}\mathrm{cos}\Theta_j\mathrm{sin}\Phi_k-V_{\mathrm{y,out}},
\end{equation}
and 
\begin{equation}
w_{\mathrm{z},ijk}=\sqrt{\frac{2E_i}{m}}\mathrm{sin}\Theta_j-V_{\mathrm{z,out}}.
\end{equation}
By replacing $f_{\mathrm{out}}(E,\Theta,\Phi)$ with $f(E,\Theta,\Phi)$ in the equations above, we calculate $N_{\mathrm{f}}$, $V_{\mathrm{f}}$, and $T_{\mathrm{f}}$, which are the density, speed, and temperature moments of the input distribution.

\section{Taylor series of the integrated distribution function}
\label{app:derivatives}

The Taylor expansion of $f(\varepsilon,\theta,\phi)$ at the center of each instrument bin $\varepsilon=E,\theta=\Theta,\phi=\Phi$, up to second-order terms, is
\begin{eqnarray}
f(\varepsilon,\theta,\phi)&\approx& f(E,\Theta,\Phi)+\frac{\partial f}{\partial\varepsilon}(E,\Theta,\Phi)(\varepsilon-E)+\frac{\partial f}{\partial\theta}(E,\Theta,\Phi)(\theta-\Theta)+\frac{\partial f}{\partial\phi}(E,\Theta,\Phi)(\phi-\Phi)\nonumber\\&+&\frac{1}{2}\left[ \frac{\partial^2f}{\partial\varepsilon^2}(E,\Theta,\Phi)(\varepsilon-E)^2 + \frac{\partial^2f}{\partial\theta^2}(E,\Theta,\Phi)(\theta-\Theta)^2+\frac{\partial^2f}{\partial\phi^2}(E,\Theta,\Phi)(\phi-\Phi)^2\right]\nonumber\\&+&\frac{\partial^2f}{\partial\varepsilon\partial\theta}(E,\Theta,\Phi)(\varepsilon-E)(\theta-\Theta)+\frac{\partial^2f}{\partial\varepsilon\partial\phi}(E,\Theta,\Phi)(\varepsilon-E)(\phi-\Phi)\nonumber\\&+&\frac{\partial^2f}{\partial\theta\partial\phi}(E,\Theta,\Phi)(\theta-\Theta)(\phi-\Phi).
\label{eq:Taylor}
\end{eqnarray}

\noindent By using the Taylor expansion of $f(\varepsilon,\theta,\Phi)$, the integral in Eq. \ref{eq:Counts} becomes
\begin{eqnarray}
C_{\mathrm{exp}}(E,\Theta,\Phi)&\approx&\Delta\tau \int\limits_{\varepsilon_{\mathrm{min}}}^{\varepsilon_{\mathrm{max}}} \int\limits_{\theta_{\mathrm{min}}}^{\theta_{\mathrm{max}}} \int\limits_{\phi_{\mathrm{min}}}^{\phi_{\mathrm{max}}}\alpha(E,\Theta,\Phi,\varepsilon,\theta,\phi)\nonumber\\&\times&\left[f(E,\Theta,\Phi)+...+\frac{\partial^2f}{\partial\theta\partial\phi}(E,\Theta,\Phi)(\theta-\Theta)(\phi-\Phi)\right]\nonumber\\&\times&\frac{2}{m^2}\varepsilon\,\mathrm{d}\varepsilon\mathrm{cos}\theta\,\mathrm{d}\theta\,\mathrm{d}\phi.
\label{eq:Counts_Taylor}
\end{eqnarray}

We now replace the linear $\varepsilon$ term with its value at the center of the bin $E$ and we write 
\begin{eqnarray}
C_{\mathrm{exp}}(E,\Theta,\Phi)&\approx&\frac{2E^2\Delta\tau}{m^2} \int\limits_{\varepsilon_{\mathrm{min}}}^{\varepsilon_{\mathrm{max}}} \int\limits_{\theta_{\mathrm{min}}}^{\theta_{\mathrm{max}}} \int\limits_{\phi_{\mathrm{min}}}^{\phi_{\mathrm{max}}}\alpha(E,\Theta,\Phi,\varepsilon,\theta,\phi)\nonumber\nonumber\\&\times&\left[f(E,\Theta,\Phi)+...+\frac{\partial^2f}{\partial\theta\partial\phi}(E,\Theta,\Phi)(\theta-\Theta)(\phi-\Phi)\right]\nonumber\\&\times&\frac{\mathrm{d}\varepsilon}{E}\mathrm{cos}\theta\,\mathrm{d}\theta\,\mathrm{d}\phi,
\label{eq:Counts_Taylor_dG}
\end{eqnarray}
which can be realized as the sum of integrals for each term of the Taylor series. The first integral, which is the integral containing the first term $f(E,\Theta,\Phi)$, is
\begin{eqnarray}
\frac{2E^2\Delta\tau}{m^2} f(E,\Theta,\Phi)&&\int\limits_{\varepsilon_{\mathrm{min}}}^{\varepsilon_{\mathrm{max}}} \int\limits_{\theta_{\mathrm{min}}}^{\theta_{\mathrm{max}}} \int\limits_{\phi_{\mathrm{min}}}^{\phi_{\mathrm{max}}}\alpha(E,\Theta,\Phi,\varepsilon,\theta,\phi)\,\frac{\mathrm{d}\varepsilon}{E}\mathrm{cos}\theta\,\mathrm{d}\theta\,\mathrm{d}\phi\nonumber\\&=&\frac{2E^2\Delta\tau G(E,\Theta,\Phi)f(E,\Theta,\Phi)}{m^2},
\label{eq:Counts_Taylor_dG2}
\end{eqnarray}
which is identical to Eq. \ref{eq:Cexp_simple}. As a result, the integrals of the higher-order terms of $f(\varepsilon,\theta,\phi)$ estimate the discrepancy between the simplified, zeroth order approach in Eq. \ref{eq:Cexp_simple}, and the exact number of counts given by Eq. \ref{eq:Counts}, which quantifies the systematic uncertainties  we investigate here. For a symmetric response function $\alpha(E,\Theta,\Phi,\varepsilon,\theta,\phi)\mathrm{cos}\theta$ around the bin center $E,\Theta,\Phi$, we get:

\begin{equation}
 <\varepsilon>=\frac{\int\limits_{\varepsilon_{\mathrm{min}}}^{\varepsilon_{\mathrm{max}}} \int\limits_{\theta_{\mathrm{min}}}^{\theta_{\mathrm{max}}} \int\limits_{\phi_{\mathrm{min}}}^{\phi_{\mathrm{max}}}\varepsilon\alpha(E,\Theta,\Phi,\varepsilon,\theta,\phi)\frac{\mathrm{d}\varepsilon}{E}\mathrm{cos}\theta\,\mathrm{d}\theta\,\mathrm{d}\phi}{\int\limits_{\varepsilon_{\mathrm{min}}}^{\varepsilon_{\mathrm{max}}} \int\limits_{\theta_{\mathrm{min}}}^{\theta_{\mathrm{max}}} \int\limits_{\phi_{\mathrm{min}}}^{\phi_{\mathrm{max}}}\alpha(E,\Theta,\Phi,\varepsilon,\theta,\phi)\frac{\mathrm{d}\varepsilon}{E}\mathrm{cos}\theta\,\mathrm{d}\theta\,\mathrm{d}\phi}=E,
\end{equation}

\begin{equation}
  <\theta>=\frac{\int\limits_{\varepsilon_{\mathrm{min}}}^{\varepsilon_{\mathrm{max}}} \int\limits_{\theta_{\mathrm{min}}}^{\theta_{\mathrm{max}}} \int\limits_{\phi_{\mathrm{min}}}^{\phi_{\mathrm{max}}}\theta\alpha(E,\Theta,\Phi,\varepsilon,\theta,\phi)\frac{\mathrm{d}\varepsilon}{E}\mathrm{cos}\theta\,\mathrm{d}\theta\,\mathrm{d}\phi}{\int\limits_{\varepsilon_{\mathrm{min}}}^{\varepsilon_{\mathrm{max}}} \int\limits_{\theta_{\mathrm{min}}}^{\theta_{\mathrm{max}}} \int\limits_{\phi_{\mathrm{min}}}^{\phi_{\mathrm{max}}}\alpha(E,\Theta,\Phi,\varepsilon,\theta,\phi)\frac{\mathrm{d}\varepsilon}{E}\mathrm{cos}\theta\,\mathrm{d}\theta\,\mathrm{d}\phi}=\Theta,
\end{equation}
and 
\begin{equation}
 <\phi>=\frac{\int\limits_{\varepsilon_{\mathrm{min}}}^{\varepsilon_{\mathrm{max}}} \int\limits_{\theta_{\mathrm{min}}}^{\theta_{\mathrm{max}}} \int\limits_{\phi_{\mathrm{min}}}^{\phi_{\mathrm{max}}}\phi\alpha(E,\Theta,\Phi,\varepsilon,\theta,\phi)\frac{\mathrm{d}\varepsilon}{E}\mathrm{cos}\theta\,\mathrm{d}\theta\,\mathrm{d}\phi}{\int\limits_{\varepsilon_{\mathrm{min}}}^{\varepsilon_{\mathrm{max}}} \int\limits_{\theta_{\mathrm{min}}}^{\theta_{\mathrm{max}}} \int\limits_{\phi_{\mathrm{min}}}^{\phi_{\mathrm{max}}}\alpha(E,\Theta,\Phi,\varepsilon,\theta,\phi)\frac{\mathrm{d}\varepsilon}{E}\mathrm{cos}\theta\,\mathrm{d}\theta\,\mathrm{d}\phi}=\Phi,
\end{equation}
and therefore, all first-order derivative terms which have $(\varepsilon-E),(\theta-\Theta)$, and $(\phi-\Phi)$ go to zero and do not contribute to the uncertainty we investigate. For this reason, the second-order derivative terms with the mixed energy, elevation, and azimuth terms, also go to zero. As a result, the systematic uncertainty of the estimated counts $\delta C_{\mathrm{exp}}$ is approximately
\begin{eqnarray}
\delta C_{\mathrm{exp}}(E,\Theta,\Phi) &\sim& \frac{E^2\Delta\tau}{m^2}  \left[\frac{\partial^2 f}{\partial\varepsilon^2}(E,\Theta,\Phi) \int\limits_{\varepsilon_{\mathrm{min}}}^{\varepsilon_{\mathrm{max}}} \int\limits_{\theta_{\mathrm{min}}}^{\theta_{\mathrm{max}}} \int\limits_{\phi_{\mathrm{min}}}^{\phi_{\mathrm{max}}}(\varepsilon-E)^2\alpha(E,\Theta,\Phi,\varepsilon,\theta,\phi)\frac{\mathrm{d}\varepsilon}{E}\mathrm{cos}\theta\,\mathrm{d}\theta\,\mathrm{d}\phi\right. \nonumber\\&+& \frac{\partial^2 f}{\partial\theta^2}(E,\Theta,\Phi)\int\limits_{\varepsilon_{\mathrm{min}}}^{\varepsilon_{\mathrm{max}}} \int\limits_{\theta_{\mathrm{min}}}^{\theta_{\mathrm{max}}} \int\limits_{\phi_{\mathrm{min}}}^{\phi_{\mathrm{max}}}(\theta-\Theta)^2\alpha(E,\Theta,\Phi,\varepsilon,\theta,\phi)\frac{\mathrm{d}\varepsilon}{E}\mathrm{cos}\theta\,\mathrm{d}\theta\,\mathrm{d}\phi \nonumber\\&+& \left.\frac{\partial^2 f}{\partial\phi^2}(E,\Theta,\Phi)\int\limits_{\varepsilon_{\mathrm{min}}}^{\varepsilon_{\mathrm{max}}} \int\limits_{\theta_{\mathrm{min}}}^{\theta_{\mathrm{max}}} \int\limits_{\phi_{\mathrm{min}}}^{\phi_{\mathrm{max}}}(\phi-\Phi)^2\alpha(E,\Theta,\Phi,\varepsilon,\theta,\phi)\frac{\mathrm{d}\varepsilon}{E}\mathrm{cos}\theta\,\mathrm{d}\theta\,\mathrm{d}\phi \right].
\label{eq:delta_Cexp}
\end{eqnarray}
We can solve the above integral, either numerically or analytically, for any input distribution and a known response function at each bin. At this point, we adopt one simplification and treat $\alpha\cos{\theta}$ (Eq. \ref{eq:response}) as it was a pure 3D-Gaussian, i.e.,

\begin{eqnarray}
\alpha(E,\Theta,\Phi,\varepsilon,\theta,\phi)\mathrm{cos}\theta&\approx&\alpha_{0}\exp\left[-\frac{\left(\frac{\varepsilon}{E}-1\right)^2}{2\left( \frac{\sigma_{E}}{E} \right)^2}\right]\exp\left[-\frac{\left( \theta - \Theta \right)^2}{2\left( \sigma_{\Theta} \right)^2}\right]\exp\left[-\frac{\left( \phi - \Phi \right)^2}{2\left( \sigma_{\Phi} \right)^2}\right]\nonumber\\&=&\alpha_{0}\exp\left[-\frac{\left(\varepsilon-E\right)^2}{2\left( \sigma_{E} \right)^2}\right]\exp\left[-\frac{\left( \theta - \Theta \right)^2}{2\left( \sigma_{\Theta} \right)^2}\right]\exp\left[-\frac{\left( \phi - \Phi \right)^2}{2\left( \sigma_{\Phi} \right)^2}\right],
\label{eq:response_gauss}
\end{eqnarray}

\noindent for which the analytical solution of Eq. \ref{eq:delta_Cexp} is straight forward. Under this approximation, we get

\begin{equation}
    \int\limits_{\varepsilon_{\mathrm{min}}}^{\varepsilon_{\mathrm{max}}} \int\limits_{\theta_{\mathrm{min}}}^{\theta_{\mathrm{max}}} \int\limits_{\phi_{\mathrm{min}}}^{\phi_{\mathrm{max}}}(\varepsilon-E)^2\alpha(E,\Theta,\Phi,\varepsilon,\theta,\phi)\frac{\mathrm{d}\varepsilon}{E}\mathrm{cos}\theta\,\mathrm{d}\theta\,\mathrm{d}\phi\approx E^{-1}\alpha_{0} (2\pi)^{3/2}\sigma_{E}^3\sigma_{\Theta}\sigma_{\Phi},
\end{equation}
\begin{equation}
    \int\limits_{\varepsilon_{\mathrm{min}}}^{\varepsilon_{\mathrm{max}}} \int\limits_{\theta_{\mathrm{min}}}^{\theta_{\mathrm{max}}} \int\limits_{\phi_{\mathrm{min}}}^{\phi_{\mathrm{max}}}(\theta-\Theta)^2\alpha(E,\Theta,\Phi,\varepsilon,\theta,\phi)\frac{\mathrm{d}\varepsilon}{E}\mathrm{cos}\theta\,\mathrm{d}\theta\,\mathrm{d}\phi\approx E^{-1}\alpha_{0} (2\pi)^{3/2}\sigma_{E}\sigma_{\Theta}^3\sigma_{\Phi},
\end{equation}
and
\begin{equation}
    \int\limits_{\varepsilon_{\mathrm{min}}}^{\varepsilon_{\mathrm{max}}} \int\limits_{\theta_{\mathrm{min}}}^{\theta_{\mathrm{max}}} \int\limits_{\phi_{\mathrm{min}}}^{\phi_{\mathrm{max}}}(\phi-\Phi)^2\alpha(E,\Theta,\Phi,\varepsilon,\theta,\phi)\frac{\mathrm{d}\varepsilon}{E}\mathrm{cos}\theta\,\mathrm{d}\theta\,\mathrm{d}\phi\approx E^{-1}\alpha_{0} (2\pi)^{3/2}\sigma_{E}\sigma_{\Theta}\sigma_{\Phi}^3,
\end{equation}
and by substituting back to Eq. \ref{eq:delta_Cexp} we get:

\begin{eqnarray}
\delta C_{\mathrm{exp}}(E,\Theta,\Phi) &=& \frac{\alpha_{0}(2\pi)^{\frac{3}{2}}\sigma_{E}\sigma_{\Theta}\sigma_{\Phi}E\Delta\tau}{m^2} \nonumber\\ &\times& \left[ \sigma_{E}^2\frac{\partial^2 f}{\partial\varepsilon^2}(E,\Theta,\Phi) +\sigma_{\Theta}^2\frac{\partial^2 f}{\partial\theta^2}(E,\Theta,\Phi)+\sigma_{\Phi}^2\frac{\partial^2f}{\partial\phi^2}(E,\Theta,\Phi)\right].
\label{eq:delta_Cexp_pre_final}
\end{eqnarray}

Under the 3D-Gaussian response approximation (Eq. \ref{eq:response_gauss}), the geometric factor of the instrument is approximately
\begin{eqnarray}
G(E,\Theta,\Phi)\approx\alpha_0(2\pi)^{\frac{3}{2}}\frac{\sigma_{E}}{E}\sigma_{\Theta}\sigma_{\Phi},
\label{eq:approx_Gfactor}
\end{eqnarray}
and thus, Eq. \ref{eq:delta_Cexp_pre_final} becomes
\begin{eqnarray}
\delta C_{\mathrm{exp}}(E,\Theta,\Phi) &=& \frac{G(E,\Theta,\Phi)E^2\Delta\tau}{m^2} \nonumber\\ &\times& \left[ \sigma_{E}^2\frac{\partial^2 f}{\partial\varepsilon^2}(E,\Theta,\Phi) +\sigma_{\Theta}^2\frac{\partial^2 f}{\partial\theta^2}(E,\Theta,\Phi)+\sigma_{\Phi}^2\frac{\partial^2f}{\partial\phi^2}(E,\Theta,\Phi)\right].
\label{eq:delta_Cexp_final}
\end{eqnarray}

The conversion from observed counts to VDF using Eq. \ref{eq:fout_simple} results in a systematic off-set in the estimation of $f_{\mathrm{out}}$, given by
\begin{equation}
\delta f(E,\Theta,\Phi)=\frac{1}{2}\left[\sigma_{E}^2\frac{\partial^2f}{\partial\varepsilon^2}(E,\Theta,\Phi)+\sigma_{\Theta}^2\frac{\partial^2f}{\partial\theta^2}(E,\Theta,\Phi)+\sigma_{\Phi}^2\frac{\partial^2f}{\partial\phi^2}(E,\Theta,\Phi)\right],
\label{eq:delta_f}
\end{equation} 
\noindent where $\sigma_{E}$, $\sigma_{\Theta}$ and $\sigma_{\Phi}$, determine the energy, elevation, and azimuth acceptance widths of the instrument bin. In general, $\sigma_{E}$, $\sigma_{\Theta}$ and $\sigma_{\Phi}$ are parameters which must be determined for each individual bin. For our concept instrument, we assume the same set of $\sigma_{E}/E$, $\sigma_{\Theta}$, and $\sigma_{\Phi}$ for all the bins, which does not affect the validity of the analysis we perform here.  
With a close look to Eq. \ref{eq:delta_f}, we understand that  the systematic uncertainty increases with increasing second-order derivatives of $f$, but also, with increasing acceptance widths (decreasing resolution) of the instrument. 

\begin{figure}
    \centering
    \includegraphics[width=0.7\linewidth]{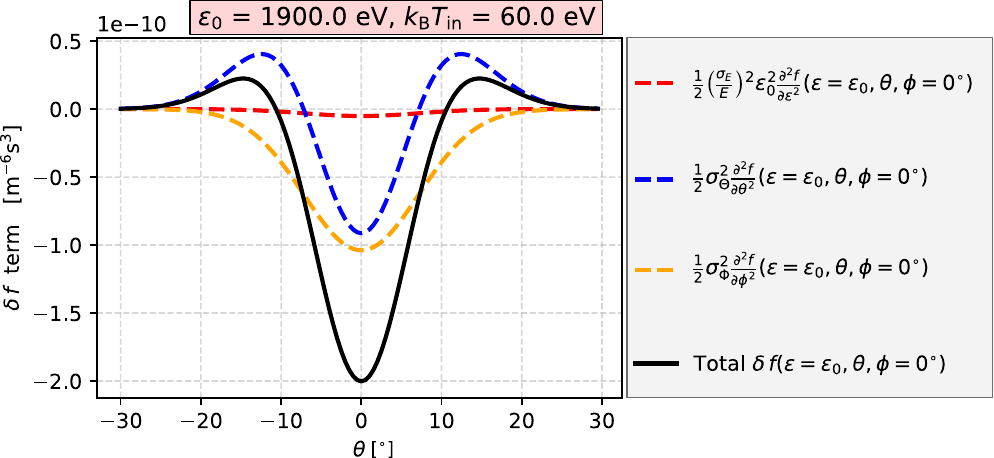}
    \caption{1D cuts of the parameter $\delta f$ (black) and its individual terms (red, blue and orange), at the bulk energy ($\varepsilon=\varepsilon_0$) and azimuth $\phi=0^{\circ}$, considering Maxwellian plasma with $N_{\mathrm{in}}=10\mathrm{\,cm}^{-3}$ $\varepsilon_0=1900$ eV and $k_{\mathrm{B}}T_{\mathrm{in}}=60$ eV. The red curve is the 1D cut of the first term of $\delta f$, which has the second-order derivative of $f$ with respect to energy. The blue curve is the 1D cut of  the second term, which has the second-order derivative of $f$ with respect to elevation, and the orange curve is the 1D cut of the third term, which has the second-order derivative of $f$ with respect to azimuth.}
    \label{fig:d2f_terms}
\end{figure}

We now evaluate $\delta f$ analytically for certain plasma properties and our concept instrument. For plasma with bulk velocity along $\theta=\phi=0^{\circ}$ direction, the Maxwell distribution in Eq. \ref{eq:maxwell_energy} becomes

\begin{equation}
f(\varepsilon,\theta,\phi)=Ae^{-\frac{\varepsilon+\varepsilon_{0}-2\sqrt{\varepsilon \varepsilon_{0}}cos\theta \cos{\phi}}{k_{\mathrm{B}}T_{\mathrm{in}}}},
\label{eq:f_en}
\end{equation}
where $A=N_{\mathrm{in}} \left(   \frac{m}{2\pi k_{\mathrm{B}}T_{\mathrm{in}}}   \right)^{\frac{3}{2}}$. Then, the partial derivative of $f$ with respect to energy is
\begin{equation}
\frac{\partial f}{\partial \varepsilon} =\frac{1}{k_{\mathrm{B}}T_{\mathrm{in}}} \left( \sqrt{\frac{\varepsilon_{0}}{\varepsilon}}\cos{\theta\cos{\phi}}-1\right)Ae^{-\frac{\varepsilon+\varepsilon_{0}-2\sqrt{\varepsilon\varepsilon_{0}}\cos{\theta}\cos{\phi}}{k_{\mathrm{B}T_{\mathrm{in}}}}}=\frac{1}{k_{\mathrm{B}}T_{\mathrm{in}}}\left( \sqrt{\frac{\varepsilon_{0}}{\varepsilon}}\cos{\theta\cos{\phi}}-1\right)f ,
\label{eq:df_de}
\end{equation} 
the partial derivative of $f$ with respect to elevation is
\begin{equation}
\frac{\partial f}{\partial \theta} =\frac{ -2\sqrt{\varepsilon \varepsilon_{0}}\sin{\theta\cos{\phi}}}{k_{\mathrm{B}}T_{\mathrm{in}}} Ae^{-\frac{\varepsilon+\varepsilon_{0}-2\sqrt{\varepsilon\varepsilon_{0}}\cos{\theta}\cos{\phi}}{k_{\mathrm{B}T_{\mathrm{in}}}}}= \frac{ -2\sqrt{\varepsilon \varepsilon_{0}}\sin{\theta\cos{\phi}}}{k_{\mathrm{B}}T_{\mathrm{in}}}f,
\label{eq:df_dth}
\end{equation} 
and the partial derivative with respect to azimuth is
\begin{equation}
\frac{\partial f}{\partial \phi} =\frac{ -2\sqrt{\varepsilon \varepsilon_{0}}\cos{\theta\sin{\phi}}}{k_{\mathrm{B}}T_{\mathrm{in}}} Ae^{-\frac{\varepsilon+\varepsilon_{0}-2\sqrt{\varepsilon\varepsilon_{0}}\cos{\theta}\cos{\phi}}{k_{\mathrm{B}T_{\mathrm{in}}}}}= \frac{ -2\sqrt{\varepsilon \varepsilon_{0}}\cos{\theta\sin{\phi}}}{k_{\mathrm{B}}T_{\mathrm{in}}}f.
\label{eq:df_dphi}
\end{equation} 
The second-order partial derivative of $f$ with respect to energy is
\begin{equation}
\frac{\partial^2 f}{\partial \varepsilon^2} =-\frac{\sqrt{\varepsilon_{0}}\cos{\theta}\cos{\phi}}{2k_{\mathrm{B}}T_{\mathrm{in}}\varepsilon^{3/2}}f+\frac{1}{k_{\mathrm{B}}T_{\mathrm{in}}}\left(\sqrt{\frac{\varepsilon_0}{\varepsilon}}\cos{\theta}\cos{\phi}-1\right)\frac{\partial f}{\partial \varepsilon} ,
\label{eq:d2f_de2}
\end{equation} 
and with the use of Eq. \ref{eq:df_de}, becomes
\begin{equation}
\frac{\partial^2 f}{\partial \varepsilon^2} =-\frac{\sqrt{\varepsilon_{0}}\cos{\theta}\cos{\phi}}{2k_{\mathrm{B}}T_{\mathrm{in}}\varepsilon^{3/2}}f+\frac{1}{k_{\mathrm{B}}^2T_{\mathrm{in}}^2}\left(\sqrt{\frac{\varepsilon_0}{\varepsilon}}\cos{\theta}\cos{\phi}-1\right)^2f .
\label{eq:d2f_de2_full}
\end{equation} 
The second-order derivative of $f$ with respect to elevation angle is
\begin{equation}
\frac{\partial^2 f}{\partial \theta^2} = -\frac{ 2\sqrt{\varepsilon \varepsilon_{0}}\cos{\theta\cos{\phi}}}{k_{\mathrm{B}}T_{\mathrm{in}}}f-\frac{ 2\sqrt{\varepsilon \varepsilon_{0}}\sin{\theta\cos{\phi}}}{k_{\mathrm{B}}T_{\mathrm{in}}}\frac{\partial f}{\partial\theta},
\end{equation} 
and with the use of Eq. \ref{eq:df_dth}, becomes
\begin{equation}
\frac{\partial^2 f}{\partial \theta^2} = -\frac{ 2\sqrt{\varepsilon \varepsilon_{0}}\cos{\theta\cos{\phi}}}{k_{\mathrm{B}}T_{\mathrm{in}}}f+\frac{ 4\varepsilon \varepsilon_{0}\sin^2{\theta\cos^2{\phi}}}{k_{\mathrm{B}}^2T_{\mathrm{in}}^2}f.
\label{eq:df_dth2_full}
\end{equation} 
Finally, the second-order derivative of $f$ with respect to the azimuth angle is
\begin{equation}
\frac{\partial^2 f}{\partial \phi^2} = -\frac{ 2\sqrt{\varepsilon \varepsilon_{0}}\cos{\theta\cos{\phi}}}{k_{\mathrm{B}}T_{\mathrm{in}}}f-\frac{ 2\sqrt{\varepsilon \varepsilon_{0}}\cos{\theta\sin{\phi}}}{k_{\mathrm{B}}T_{\mathrm{in}}}\frac{\partial f}{\partial\phi},
\end{equation} 
and with the use of Eq. \ref{eq:df_dphi}, becomes
\begin{equation}
\frac{\partial^2 f}{\partial \phi^2} = -\frac{ 2\sqrt{\varepsilon \varepsilon_{0}}\cos{\theta\cos{\phi}}}{k_{\mathrm{B}}T_{\mathrm{in}}}f+\frac{ 4\varepsilon \varepsilon_{0}\cos^2{\theta\sin^2{\phi}}}{k_{\mathrm{B}}^2T_{\mathrm{in}}^2}f.
\label{eq:df_dphi2_full}
\end{equation} 
\noindent  We now evaluate  Eq. \ref{eq:delta_f}, using the above expressions for the second-order plasma derivatives. Figure \ref{fig:d2f_terms} shows 1D cuts of the individual terms of $\delta f$ at $\varepsilon=\varepsilon_0$ and $\phi=0^{\circ}$, calculated for our concept instrument acceptance widths and a Maxwellian $f$ with density $N_{\mathrm{in}}=10\,\mathrm{cm^{-3}}$, bulk energy $\varepsilon_0=1.9\,\mathrm{keV}$ (velocity along $\theta=\phi=0^{\circ}$), and temperature $k_{\mathrm{B}}T_{\mathrm{in}}=60\,\mathrm{eV}$. The first term, which describes the error due to the unresolved changes of $f$ over $\varepsilon$ within each bin, is the smallest. The second and third terms are the dominant terms, indicating that the unresolved changes of $f$ over elevation and azimuth within the instrument bins, contribute most to the errors we investigate in this study.

\begin{figure}
    \centering
    \includegraphics[width=0.8\linewidth]{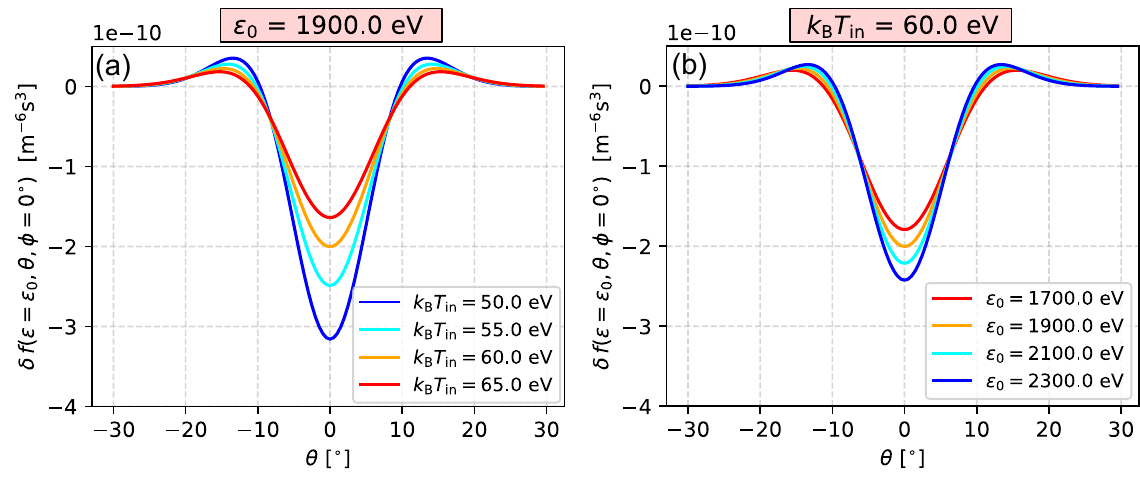}
    \caption{The $\delta f$ function and its dependence on the plasma bulk energy and temperature. \textbf{(a)} 1D cuts of $\delta f$ at the bulk energy ($\varepsilon=\varepsilon_0$) and azimuth $\phi=0^{\circ}$, for four  Maxwellian distributions with the same bulk energy $\varepsilon_0=1.9$ keV, but different input temperatures;  $k_{\mathrm{B}}T_{\mathrm{in}}=50$ eV (blue), $k_{\mathrm{B}}T_{\mathrm{in}}$=55 eV (cyan), $k_{\mathrm{B}}T_{\mathrm{in}}$=60 eV (orange), and $k_{\mathrm{B}}T_{\mathrm{in}}=65$ eV (red). \textbf{(b)} 1D cuts of $\delta f$ at the bulk energy ($\varepsilon=\varepsilon_0$) and azimuth $\phi=0^{\circ}$, for four Maxwellian distributions with the same temperature $k_{\mathrm{B}}T_{\mathrm{in}}=60$ eV, but different bulk energies;  $\varepsilon_0=1700$ eV (red), $\varepsilon_0=$ 1900 eV (orange), $\varepsilon_0=2100$ eV (cyan), and $\varepsilon_0$=2300 eV(blue). All distributions have the same density, $N_{\mathrm{in}}=10\mathrm{\,cm}^{-3}$.}
    \label{fig:d2f_vs_e0_vs_T}
\end{figure}

We also examine the behavior of $\delta f$ for different input speeds (bulk energies) and temperatures. Figure \ref{fig:d2f_vs_e0_vs_T}\textbf{(a)} shows 1D cuts of $\delta f$ at the bulk energy ($\varepsilon=\varepsilon_0$) and azimuth $\phi=0^{\circ}$, for four  Maxwellian distributions with the same density, $N_{\mathrm{in}}=10\mathrm{\,cm}^{-3}$, the same bulk energy, $\varepsilon_0=1.9$ keV, but different input temperatures. The blue curve is the calculation for $k_{\mathrm{B}}T_{\mathrm{in}}=50$ eV, the cyan curve for $k_{\mathrm{B}}T_{\mathrm{in}}=55$ eV, the orange curve for $k_{\mathrm{B}}T_{\mathrm{in}}=60$ eV, and the red curve is for $k_{\mathrm{B}}T_{\mathrm{in}}=65$ eV. Figure \ref{fig:d2f_vs_e0_vs_T}\textbf{(b)} shows the same cuts for four Maxwellian distributions with the same density, $N_{\mathrm{in}}=\mathrm{\,cm}^{-3}$, same temperature, $k_{\mathrm{B}}T_{\mathrm{in}}$ = 60 eV, but different bulk energies. The red curve is for $\varepsilon_0$ = 1700 eV, the orange curve is for $\varepsilon_0$ = 1900 eV, the cyan curve for $\varepsilon_0$ = 2100 eV, and the blue curve is for $\varepsilon_0$ = 2300 eV. We see that $\delta f$ is always negative at the core, and its minimum value decreases with increasing speed and/or decreasing temperature. For colder and/or faster protons, $\delta f$ becomes positive for smaller absolute $\theta$ values, and exhibits local maxima that are greater than $\delta f$ functions for slower and/or hotter protons. Although the derivatives of $f$ vary within individual bins, for the certain example we show in Fig. \ref{fig:derivatives_sketch}, $\delta f(E,\Theta,\Phi)$ evaluated at the center of the bins captures the uncertainties accurately. However, for colder and faster distributions, we recommend evaluating higher order derivatives of $f$ as well, in order to capture the uncertainties with the analytical expression accurately.

\bibliography{references}

\end{document}